\documentclass{aastex}
\usepackage{natbib}
\shorttitle{Conditions for Alfv\'{e}nic oscillations to heat the chromosphere.}
\shortauthors{Goodman} 

\begin{document}

\title{Conditions for Photospherically Driven Alfv\'{e}nic Oscillations to Heat the Solar Chromosphere by Pedersen Current Dissipation}  
\author{Michael L. Goodman}   
\affil{Advanced Technologies Group, West Virginia High Technology Consortium Foundation\\1000 Galliher Drive, Fairmont, WV 26554}
\email{mgoodman@wvhtf.org}

\begin{abstract}
An MHD model that includes a complete electrical conductivity tensor is used to estimate conditions for photospherically driven, linear, non-plane Alfv\'{e}nic oscillations extending from the photosphere to the lower corona to drive a chromospheric heating rate due to Pedersen current dissipation that is comparable to the observed net chromospheric
radiative loss of  $\sim 10^7$ ergs-cm$^{-2}$-sec$^{-1}$. The heating rates due to electron current dissipation in the photosphere and corona are also computed. The wave amplitudes are computed self-consistently as functions of an inhomogeneous background atmosphere. The effects of the conductivity tensor are resolved numerically using a resolution of 3.33 m. The oscillations drive a chromospheric heating flux $F_{Ch} \sim 10^7-10^8$ ergs-cm$^{-2}$-sec$^{-1}$ at frequencies $\nu \sim 10^2 - 10^3$ mHz for background magnetic field strengths $B \gtrsim 700$ G, and magnetic field perturbation amplitudes $\sim 0.01-0.1 \; B$.
The total resistive heating flux increases with $\nu$. Most heating occurs in the photosphere. Thermalization of Poynting flux in the photosphere due to electron current dissipation regulates the Poynting flux into the chromosphere, limiting $F_{Ch}$. $F_{Ch}$ initially increases with $\nu$, reaches a maximum, and then decreases with increasing $\nu$ due to increasing electron current dissipation in the photosphere. The resolution needed to resolve the oscillations increases from $\sim 10$ m in the photosphere to $\sim 10$ km in the upper chromosphere, and is $\propto \nu^{-1/2}$. Estimates suggest these oscillations are normal modes of photospheric flux tubes with diameters $\sim 10-20$ km, excited by magnetic reconnection in current sheets with thicknesses $\sim 0.1$ km.  
\end{abstract}

\section*{1. Introduction}

The chromosphere is weakly ionized and strongly magnetized in regions with photospheric magnetic field strengths $\gtrsim 10^2$ G. This distinguishes it from the underlying weakly ionized, weakly magnetized photosphere, and the overlying strongly ionized, strongly magnetized corona. The combination of weak ionization and strong magnetization implies that the main MHD resistive heating mechanism is the dissipation of ion (mainly proton) Pedersen currents. This dissipation is characterized by a Pedersen resistivity orders of magnitude greater than the Spitzer resistivity. This heating mechanism is not effective in the photosphere due to weak magnetization, or in the corona due to strong ionization.
The question is whether there are drivers that generate electric fields in the chromosphere strong enough to drive a
resistive heating rate due to Pedersen current dissipation that is comparable to the net radiative loss (NRL) of the chromosphere, which is $\sim 10^7$ ergs-cm$^{-2}$-sec$^{-1}$. Alfv\'{e}n waves are a potential driver.

Observations using the Solar Optical Telescope (SOT) aboard the Hinode satellite suggest the presence of 
Alfv\'{e}n waves throughout the upper chromosphere and lower transition region (TR) (De Pontieu et al. 2007 a; McIntosh,
De Pontieu \& Tarbell 2008). The temporal resolution of these observations is $\sim 5$ sec.
The observed waves occur in spicules with diameters $\sim 200$ km that pervade this region of the atmosphere, and appear as vertical or parabolic jets of heated plasma along magnetic field lines from the chromosphere to the TR (De Pontieu et al. 2007 b). The existence of Alfv\'{e}n waves is inferred from observations of constant velocity, transverse displacements of spicules at amplitudes $\delta V \sim 10 - 30$ km-sec$^{-1}$. These authors propose that the absence of oscillations of the velocity amplitude expected of Alfv\'{e}n waves in the observations is due to the
fact that the inferred wave periods of $\sim 100 - 500$ seconds are greater than the observed spicule lifetimes of $\sim
10 - 300$ seconds, with most lifetimes $< 100$ seconds. The inferred wavelengths of the Alfv\'{e}n waves are $\gtrsim 4
\times 10^3$ km. The estimated range of the Alfv\'{e}n speed $V_A$ at the location of the observed transverse velocities is $\sim 50 - 200$ km-sec$^{-1}$. Then the frequency $\nu$ of these waves is $\lesssim 50$ mHz. The observed waves satisfy the condition for linear waves. This is shown as follows. The linearized magnetohydrodynamic (MHD) equations for
Alfv\'{e}n waves imply that $\delta B/B_0 = \delta V/V_A$, where $B_0$ and $\delta B$ are the background magnetic field strength, and amplitude of the magnetic field perturbation. The condition for linearity is $\delta B/B_0 <1$. Since $\delta V \sim 10 - 30$ km-sec$^{-1}$, it follows that $\delta V/V_A \sim 0.05 - 0.6$, so the waves are linear. 

The penetration of Alfv\'{e}n waves into the corona, and the transformation of their energy into thermal and center of mass (CM) kinetic energy is of longstanding interest. MHD linear wave theory suggests these waves exist in the corona as a basic MHD wave mode, and observations confirm their existence (e.g. Tomczyk \& McIntosh 2009).

These observations give new importance to the old question of whether linear Alfv\'{e}n waves can be an important driver of chromospheric heating, and source of Poynting flux into the corona.

The purpose of this paper is to estimate conditions under which linear, non-plane Alfv\'{e}nic oscillations that extend from the photosphere to the lower corona, and that are driven by boundary conditions at the photosphere generate an electric field that drives chromospheric heating rates $\gtrsim 10^7$ ergs-cm$^{-2}$-sec$^{-1}$ by Pedersen current dissipation. The resistive heating rates
due to electron current dissipation in the photosphere and corona are also computed. The height dependent wave amplitudes are computed self-consistently as functions of an inhomogeneous background atmosphere given by model CM of Fontenla, Avrett \& Loeser (E.H. Avrett \& R. Loeser 2001, private communication, henceforth FAL). It is essentially the model
described in Fontenla, Avertt \& Loeser (2002). The FAL background state profiles are shown in figure 1.
The following results are obtained. The oscillations drive significant chromospheric heating at frequencies $\nu \sim 10^2 - 10^3$ mHz in the presence of background magnetic field strengths $B \gtrsim 700$ G, assuming magnetic field perturbation amplitudes $\sim 0.01-0.1 \; B$. The thermalization of Poynting flux in the photosphere due to electron current dissipation can play a major role in regulating the Poynting flux that flows into the chromosphere, and hence in limiting the chromospheric heating flux. This flux increases with frequency until electron current dissipation in the photosphere becomes so large that the chromospheric heating flux decreases with any further increase in $\nu$. The spatial resolution needed to resolve these oscillations increases from $\sim 10$ m in the photosphere to $\sim 10$ km in the upper chromosphere. The estimated coronal heating flux up to a height $z= 19000$ km, which is the maximum height in the model, is $\sim 10^2-10^3$ ergs-cm$^{-2}$-sec$^{-1}$. This is too small by several orders of magnitude to balance coronal energy losses, and provide the thermal component of the energy for accelerating the solar wind. The Poynting flux at $z=19000$ km due to these oscillations is $\sim 3 \times 10^6 - 2 \times 10^7$ ergs-cm$^{-2}$-sec$^{-1}$. This is sufficient to provide the required energy input into the corona in active regions and coronal holes, which is $\sim 10^7$ and $10^6$ ergs-cm$^{-2}$-sec$^{-1}$, respectively, and exceeds the required energy input into the corona in quiet regions, which is $\sim 3 \times 10^5$ ergs-cm$^{-2}$-sec$^{-1}$, consistent with the weaker average magnetic field strengths in quiet regions (Withbroe \& Noyes 1977; Jordan 1981). Estimates presented in 
\S 7 support the proposition that these oscillations are normal modes of photospheric flux tubes with diameters $\sim 10-20$ km, excited by magnetic reconnection in current sheets with thicknesses $\sim 0.1$ km. These diameters are
consistent with the highest resolution observations of G-band bright points in the photosphere, discussed later in this
section, that place an upper bound of $0.09^{''}$ on the flux tube diameter. The coupling of the vertical wavelength to the radial boundary condition within the flux tube, combined with the diameter of the flux tube that
gives rise to a spectrum of Alfv\'{e}n waves with frequencies $\sim 1$ Hz.  

The first modeling work to present a compelling argument that linear Alfv\'{e}n wave driven Pedersen current dissipation can be a major source of chromospheric heating, and that the damping rate increases rapidly with height due to the increasing magnetization of the plasma appears to be that of De Pontieu \& Haerendel (1998), and De Pontieu, Martens \& Hudson (2001). The frequency range of these waves $\sim 10^2 - 10^3$ mHz.
MHD simulations of driven, low amplitude Alfv\'{e}n wave damping by Leake, Arber \& Khodachenko (2005) support the main results of De Pontieu \& Haerendel (1998) and De Pontieu, Martens 
\& Hudson (2001). Osterbrock (1961), following similar work by Piddington (1956), presents an analysis of damping rates
of linear MHD waves in the chromosphere due to charged particle-neutral collisions, and a scalar viscosity, and concludes that damping of these waves is not important for chromospheric heating. However, due to observational limitations at that time, Osterbrock (1961) uses magnetic field strengths orders of magnitude smaller than those now known to exist, causing the heating rate due to charged-particle neutral collisions to be under estimated by orders of magnitude. Khodachenko, Arber, Rucker \& Hanslmeier (2004) model the relative importance of chromospheric heating by dissipation of linear MHD waves due to Pedersen current and viscous dissipation, and conclude that they are of comparable importance. Kazeminezhad \& Goodman (2006, henceforth KG06) present nonlinear MHD simulations of the dissipation of Alfv\'{e}nic oscillations and wave trains due to Pedersen current dissipation in the chromosphere. Those simulations suggest these disturbances are damped out over a distance from their point of generation $\sim$ their wavelength, and might be important for localized chromospheric heating. A detailed discussion of this earlier work is deferred to \S 9 of this paper where a comparison
is made between the approaches, approximations, and results of these models in the context of the model and results presented here.  

The models of De Pontieu \& Haerendel (1998), De Pontieu, Martens \& Hudson (2001), Leake, Arber \& Khodachenko (2005),
and KG06 each have some combination of the following limitations: (1) The linear perturbation amplitudes are not consistent with the inhomogeneous background (BG) state. (2) None of the models include a horizontal component of the BG magnetic field, although the use of a $B(z)$ in these models to compute the conductivity tensor is consistent with pure Alfv\'{e}n waves in 1D only if the BG field has a horizontal component. If the BG field does not have a horizontal component, it must be a constant vertical field in order to be consistent with Alfv\'{e}n waves. If the BG field has a horizontal component then magnetoacoustic waves are coupled into the model. (3) A height dependent BG magnetic field in general implies resistive heating in the BG state. This is not discussed in any of these papers. (4) Some models do not include the lower chromosphere or photosphere. (5) Some models omit the Hall terms, or artificially amplify them. (6) None of the simulations use a resolution sufficient to fully resolve the resistive heating and Hall terms. The model presented here is simple, but does not have these limitations. 

Results from a model preliminary to the one developed here are presented in Goodman \& Kazeminezhad (2010 b). That model also considers photospherically driven, linear Alfv\'{e}nic oscillations in an inhomogeneous background atmosphere, but assumes the atmosphere is spatially slowly varying to allow for an analytic solution for the wave amplitudes. The model presented here includes the full effects of the inhomogeneous background atmosphere on the perturbation amplitudes. The
results differ significantly from those in Goodman \& Kazeminezhad (2010 b), and are assumed to be more realistic.

Using a spatial resolution sufficient to resolve resistive and other diffusive transport processes is important for the following reason. These kinetic processes directly determine resistive and viscous heating rates and diffusive thermal energy fluxes, and indirectly influence the compressive heating rate. Furthermore, radiative transition
rates, and hence the radiation field are exponentially sensitive functions of temperature through the Planck function, and diffusive transport processes play a large role in determining the temperature. It follows that the radiation field
is a sensitive function of the transport processes, so accurately computing transport processes is necessary for
accurately predicting the radiation field. Accurately computing these processes is a challenge for multidimensional MHD simulations, which are currently limited to using spatial resolutions orders of magnitude larger than what is needed, in order that the runtime not be impractically large. 

Detailed descriptions of the roles of electrons, protons, heavy ions, the neutral gas, and the magnetic field in determining the values and height dependence of the Hall, Pedersen, and Spitzer conductivities in the anisotropic electrical conductivity tensor, and how they cause the transition of the resistive heating mechanism from electron current dissipation in the photosphere governed by the Spitzer resistivity, to proton and heavy ion Pedersen current dissipation in the chromosphere governed by the Pedersen resistivity, and back to electron current dissipation in the corona, along with an analysis of the height dependence of the efficiency of resistive heating in converting electromagnetic energy into
thermal energy are presented in Goodman (2000, 2001, 2004 a,b), KG06, and Goodman \& Kazeminezhad (2010 a). The models in these papers also determine conditions under which significant chromospheric heating by Pedersen current dissipation may be driven by linear slow magnetoacoustic waves $(\nu \sim 0.8 - 3.5 \; \mbox{mHz})$, steady bulk flow $\perp {\bf B}$, nonlinear Alfv\'{e}n waves, and fast magnetoacoustic shock waves. 

This paper focuses on Alfv\'{e}n wave driven heating. As just indicated, there may be several drivers. Previous modeling 
suggests the convection electric field generated by steady CM flow of weakly ionized gas $\perp {\bf B}$ is 
an important driver, and so far seems to be the only non-wave driver.\footnote{Here steady flow means flow that is slowly varying relative to the periods found necessary for significant heating in wave driven models such as the one
presented here, with the caveat that some types of wave driving can be dominated by the wave generated convection electric field (Goodman 2000).} Convection driven heating involves the conversion of CM kinetic energy into thermal energy.\footnote{This is the operating principle of MHD power generators in which weakly ionized plasma is driven across magnetic field lines, generating a convection electric field that drives current (Rosa 1987).} At present the relative importance of wave and non-wave drivers is not known. Theoretical evidence for the importance of convection driven heating, and its possible connection with observations of small scale magnetic structures in the photosphere are briefly summarized in the remainder of this section. The 2.5 D MHD models of Goodman (1997a,b), which extend the models in Goodman (1995, 1996), describe middle chromospheric heating by Pedersen current dissipation when the current is driven by a steady state convection electric field. Since these models are restricted to the middle chromosphere the density is $\lesssim 10^{14}$ cm$^{-3}$. The main conclusion to be drawn from these models is that significant middle chromospheric heating can occur in horizontally localized, closed magnetic structures with characteristic horizontal scales $\sim 10^2 - 10^3$ km. These scales are consistent with the horizontal scales $\sim 10-10^2$ km predicted by 2.5 D MHD models of horizontally localized magnetic flux tubes and closed magnetic structures based in the photosphere and lower chromosphere 
(Goodman 2000, 2004b). The scales are consistent because the characteristic horizontal scale of the magnetic field is expected to increase with increasing height due to decreasing gas pressure. These estimates of the horizontal dimensions of strongly heated kilogauss and hectogauss magnetic structures in the photosphere and chromosphere, respectively, are
consistent with the dimensions of similar structures in the photosphere inferred from recent Hinode SOT/SP and Swedish
Solar Telescope (SST) observations of network and internetwork (IN) regions (S\'{a}nchez Almeida, Bonet, Viticchi\'{e} \& 
Del Moro 2010; Viticchi\'{e}, S\'{a}nchez Almeida, Del Moro \& Berrilli 2011). These two sets of observations have
respective spatial resolutions $\sim 0.32^{''}$ and $0.1^{''}$. The inferred range of field strengths is $\sim 1258 - 
1644$ G, with the stronger fields concentrated in the network. The inferred filling factors of these structures is
$\sim 4.5 \% \;(2.3 \% \; \mbox{(IN)} + 2.2 \% \; \mbox{(network)})$ for SOT/SP, and $\sim 2.97 \% \; (\sim 0.77 \% \; \mbox{(IN)} + 2.2 \% \; \mbox{(network)})$ for SST. Hectogauss fields are also detected with larger filling factors. A conclusion of the analysis of these and earlier observations (S\'{a}nchez Almeida et al. 2004; de Wijn, Rutten, Haverkamp \& S\"{u}tterlin 2005; Bovelet \& Wiehr 2008) is that kilogauss magnetic fields are necessary to explain the presence of G-band bright points in intergranular lanes in quiet Sun network and IN. There is an observational upper limit 
$\sim 0.09-1.0^{''}$ on the diameter of the magnetic field concentrations that give rise to G-band bright points, and the smallest G-band structures are not resolved (Bovelet \& Wiehr 2008; Viticchi\'{e}, Del Moro, Criscuoli \& Berrilli 2010).

As spatial resolution has increased, stronger fields have been detected at smaller scales, and the inferred photospheric filling factors of kilogauss and hectogauss magnetic structures in the network and IN has increased. This provides increasing observational support for MHD based mechanisms of chromospheric heating in the network and IN. Regarding heating of the IN, where acoustic shocks of the type predicted by Carlsson \& Stein (1997; also see Carlsson \& Stein
1992, 1994, 1995, 2002) are clearly observed (e.g. W\"{o}ger, Wedemeyer-B\"{o}hm, Uitenbroek \& Rimmele 2009; Vecchio, Cauzzi \& Reardon 2009), a sequence of observational studies leads to the conclusion that magnetic field concentrations in the IN strongly suppress these shocks, and that the chromospheric NRL in the IN is strongly affected, and possibly dominated by MHD processes (Judge \& Carpenter 1998; Judge, Tarbell \& Wilhelm 2001; Judge, Carlsson \& Stein 2003; Judge, Saar, Carlsson \& Ayres 2004; Vecchio, Cauzzi \& Reardon 2009). 

The observations cited above imply that chromospheric heating by Pedersen current dissipation must be given serious
consideration as a major heating mechanism given that it is the main resistive heating mechanism of the chromosphere,
that it operates throughout the chromosphere, and that it can be driven by several MHD processes.

\section*{2. Model Equations}

The 1.5 D MHD model equations are the linearized form of equations (6)-(12) in KG06.
These equations consist of the mass, momentum, and energy conservation equations, Faraday's law, and the ideal gas
equation of state. The equations include an Ohm's law with a complete electrical conductivity tensor for a three
component plasma of electrons, one species of neutral atoms, and one species of singly ionized atoms used to represent
the solar atmosphere (Goodman 2004a). The conductivity tensor is based on one derived by Mitchner \& Kruger (1973, Chapter 4 , Sec. 8). The neutral species represents HI and HeI. The singly charged ion species represents protons and
singly charged heavier ions (e.g. HeII, FeII, CaII, MgII, SiII, OII). 

Here these equations are linearized about a 1 D background state defined by a constant vertical magnetic field $B_z$,
and the FAL pressure, temperature, and particle density profiles. There is no flow in the
background state. The variable $z$ measures height above the photosphere at $z=0$. Cartesian coordinates $x,y,z$ are used.

The linear perturbation is assumed to depend only on $z$ and $t$, and to oscillate with an $\exp(i \omega t)$ time
dependence. Here $\omega$ is a real driving frequency. It is an input to the model. The height dependence of the perturbation in the inhomogeneous background atmosphere is determined by solving a set of differential equations given
boundary conditions at $z=0$. 

The frequency $\nu(=\omega/2\pi)$ is not arbitrary. It must be sufficiently small so the Ohm's law is valid. This is briefly explained as follows under the assumption of the validity of a multi-fluid MHD description of the plasma, with details
given by Mitchner \& Kruger (1973). The Ohm's law is a simplified form of the generalized Ohm's law derived by Mitchner \&
Kruger (1973) by forming a linear combination of the electron and ion momentum equations in their three fluid model. The simplification of the generalized Ohm's law carried out by Mitchner \& Kruger (1973) is based on four assumptions, each of
which places an approximate lower bound on the characteristic time scale for a change in the macroscopic state of the plasma. For the case considered here, this characteristic time scale is $\nu^{-1}$, so the assumptions place upper bounds on $\nu$. For the FAL background state, and for $0 \leq z(\mbox{km}) \leq 2100$ these assumptions require $\nu \ll 720$ Hz. The frequencies used in this paper satisfy $\nu \lesssim 3.4$ Hz, so this requirement is satisfied. None of the four assumptions require $\nu \ll \nu_{ni}$, where $\nu_{ni}$ is the neutral-ion collision frequency. However,
for the FAL background state, and for $0 \leq z(\mbox{km}) \leq 2100$, it is found that $\nu_{ni} \gtrsim 280$ Hz, so $\nu \ll \nu_{ni}$. This means that the macroscopic electromagnetic force generated by the wave, which acts directly on the ions, is strongly coupled to the neutral gas through collisions. Provided $\nu \ll \nu_{in}+\nu_{ni}$, which is one of the four assumptions used to reduce the generalized Ohm's law, the reduced Ohm's law remains valid for $\nu \gtrsim 
\nu_{ni}$, which may correspond, for example, to a sufficiently weakly ionized gas, but in that case the electromagnetic
force plays a minor role in determining the neutral gas dynamics.

For the assumed background state the linearized equations de-couple into two groups. One group determines the perturbations for mass density $\rho$, pressure $p$, temperature $T$, and vertical velocity $V_z$. This set of perturbations determines purely acoustic modes. The second group determines the perturbations for the horizontal magnetic field and velocity components $B_x, B_y, V_x$, and $V_y$. This set of perturbations defines Alfv\'{e}n modes. Only the
Alfv\'{e}n modes are considered here. The acoustic and Alfv\'{e}n modes are coupled if the background state has a nonzero horizontal magnetic field, resulting in magnetoacoustic modes that might be important drivers of resistive heating.

Let $f_1(z,t)$ be the real perturbation of any quantity $f$ at frequency $\omega$. Then $f_1(z,t)= Re(f_{1\omega}(z) \exp(i \omega t))$. Here $f_{1\omega}$ is the complex amplitude at frequency $\omega$, and $Re$ denotes the
real part. Then $f_1=f_{1 \omega R} \cos \omega t - f_{1 \omega I} \sin \omega t$. Here $f_{1 \omega R}$ and 
$f_{1 \omega I}$ are the real and imaginary parts of $f_{1 \omega}$. The time average of the product of any two
quantities $f_1(z,t)$ and $g_1(z,t)$ over a period $T=2\pi/\omega$ is 
\begin{equation}
\langle f_1 g_1 \rangle = (f_{1 \omega R} g_{1 \omega R} + f_{1 \omega I} g_{1 \omega I})/2 \equiv f_{1\omega} \cdot g_{1\omega}/2. \label{100} 
\end{equation}
This expression is used in to compute time averages. Let $f_0(z)$ denote the background profile of $f$. 

The equations for the Alfv\'{e}n modes are as follows.

The $x$ and $y$ components of the momentum conservation equation are
\begin{eqnarray}
i \omega \rho_0(z) V_{x1 \omega}(z) &=& \frac{B_z}{4 \pi} B_{x1 \omega}^\prime(z) \label{4} \\
i \omega \rho_0(z) V_{y1 \omega}(z) &=& \frac{B_z}{4 \pi} B_{y1 \omega}^\prime(z). \label{5}
\end{eqnarray}
Here the prime denotes $d/dz$.

Using these equations, the $x$ and $y$ components of Faraday's law may be written as 
\begin{equation}
\left(\left(D_P(z) - \frac{i V_A^2(z)}{\omega}\right) B_{x1 \omega}^\prime(z) - D_H(z) 
B_{y1 \omega}^\prime(z)\right)^\prime =  i \omega B_{x1\omega}(z) \label{6}
\end{equation}
and
\begin{equation}
\left(\left(D_P(z) - \frac{i V_A^2(z)}{\omega}\right) B_{y1 \omega}^\prime(z) + D_H(z) 
B_{x1 \omega}^\prime(z)\right)^\prime =  i \omega B_{y1\omega}(z). \label{7}
\end{equation}
Here $V_A (=B_z/(4 \pi \rho_0(z))^{1/2}),D_H$, and $D_P$ are the Alfv\'{e}n speed, and the Hall and Pedersen diffusivities. The diffusivities are defined in terms of the Hall and Pedersen resistivities $\eta_H$ and $\eta_P$ by $D_H = - c^2 
\eta_H/4 \pi$ and $D_P = c^2 \eta_P/4 \pi$. The Hall and Pedersen resistivities are defined as follows.
\begin{eqnarray}
\eta_H &=& \frac{B_z}{e c n_e(z)}  \label{8} \\
\eta_P &=& (1+\Gamma(z)) \eta_\parallel(z)  \label{9} \\
\Gamma & = & \left(\frac{\rho_n(z)}{\rho(z)}\right)^2 M_e(z) M_i(z)  \label{10}
\end{eqnarray}
Here $n_e, e, \eta_\parallel, \rho_n$, and $\rho$ are the electron number density and charge magnitude, resistivity for current flow parallel to the magnetic field (i.e. Spitzer resistivity), and the neutral and total mass densities.
The only way multi-fluid effects enter the model is through the charged and neutral particle densities that determine 
$\eta_H, \eta_P$, and $\eta_\parallel$, which also depend on temperature. These densities and this temperature are
assumed to be given by their FAL profiles. It remains to specify the magnitude of the constant background field $B_z$ to
determine the resistivities.

Given $B_z, \omega$, and appropriate boundary conditions, equations (\ref{6}) and (\ref{7}) determine $B_{x1 \omega}(z)$ and 
$B_{y1 \omega}(z)$. Here the appropriate boundary conditions are chosen to be the values of $B_{x1 \omega}(0), 
B_{x1 \omega}^\prime(0), B_{y1 \omega}(0)$, and $B_{y1 \omega}^\prime(0)$. Then equations (\ref{4}) and 
(\ref{5}) determine $V_{x1 \omega}$ and $V_{y1 \omega}$. 

\section*{3. Determination of Boundary Conditions at the Photosphere}

Although the values of $B_{x1 \omega}(0), B_{x1 \omega}^\prime(0), B_{y1 \omega}(0)$, and $B_{y1 \omega}^\prime(0)$ may be chosen arbitrarily, here it is assumed these values correspond to a linear, plane Alfv\'{e}n wave at $z=0$.
The corresponding Alfv\'{e}nic oscillation in the overlying atmosphere is modified from its plane wave form by the inhomogeneity of the atmosphere. The wave at $z=0$ is assumed to have an $\exp i(\omega t + k(\omega) z)$ dependence, with 
$\omega$ real and $Re(k)<0$. The choice $Re(k)<0$ implies the phase velocity at $z=0$ is directed upward. The corresponding solutions have a vertical Poynting flux $S_z(0) >0$, corresponding to electromagnetic energy flowing through the photosphere into the overlying atmosphere where some of it is resistively dissipated. The corresponding solutions have $S_z(z) >0$ for $z \geq 0$. For the solutions considered here, $0 \leq z \leq 1.9 \times 10^4$ km, so they extend from the photosphere into the lower corona. 

There are also solutions with $Re(k)<0$. They correspond to a downward phase velocity at $z=0$, and $S_z(z) <0$ for 
$z \geq 0$. These solutions correspond to electromagnetic energy flowing from the corona into the underlying atmosphere. These solutions might describe a process in which a downward Poynting flux is generated in the corona, perhaps by magnetic reconnection.

The equations for linear, plane waves in the presence of the anisotropic electrical conductivity tensor embodied in the
Ohm's law used here are given in Sec. 3 of KG06. The replacement $\omega \rightarrow - \omega$ must be made in those equations so they apply to the time dependence assumed here. Those equations are used to describe the assumed plane 
Alfv\'{e}n wave at $z=0$. It is shown in this section that they allow the solution to the model to be determined by specifying only $B_{x1\omega}(0)$.

The Alfv\'{e}nic oscillations considered here are driven in that $\omega$ is real so there is no damping in time. This means there is a monochromatic source of electromagnetic energy, presumably photospheric convection interacting with the magnetic field, that continually drives a Poynting flux into the upper atmosphere, assuming boundary conditions at the
photosphere corresponding to upward propagating waves. 

\subsection*{3.1. The Dispersion Relation $k(\omega)$}

The dispersion relation for Alfv\'{e}n waves given by equation (48) in KG06 is used to express $k$ as a function of 
$\omega$ and the background state at $z=0$. That dispersion relation gives the following four solutions for $k(\omega)$.
\begin{eqnarray}
k &=& \pm \frac{\omega}{V_{A0}} \left[ \frac{\left(1 \pm \frac{\omega}{{\bar{\omega}}_{pz}}\right) -i \left(\frac{\omega D_P}{V_{A0}^2}\right)}
{\left(1 \pm \frac{\omega}{{\bar{\omega}}_{pz}}\right)^2 + \left(\frac{\omega D_P}{V_{A0}^2}\right)^2 }\right]^{1/2} \nonumber \\
&= & \pm \frac{\omega}{V_{A0}} \left[ \frac{\left(1 \pm \frac{\omega |D_H|}{ V_{A0}^2 }\right) -i \left(\frac{\omega D_P}{V_{A0}^2}\right)}
{\left(1 \pm \frac{\omega |D_H|}{V_{A0}^2}\right)^2 + \left(\frac{\omega D_P}{V_{A0}^2}\right)^2 }\right]^{1/2} \equiv 
\pm k_\pm. \label{21}
\end{eqnarray}

Here ${\bar{\omega}}_{pz}=\omega_{pz} n_{e0}/n_0$ where $\omega_{pz}$ is the proton cyclotron frequency computed using
the constant magnetic field strength $B_z$. The quantities $n_{e0}$ and $n_0$ are the background electron and total
number densities where $n_0 \equiv \rho_0/m_p$, $\rho_0$ is the total background mass density, and $m_p$ is the proton
mass. The $\pm$ signs inside the bracket distinguish between the two wave modes that arise from the splitting of the 
Alfv\'{e}n mode due to the Hall effect. This splitting $\rightarrow 0$ as $\omega/{\bar{\omega}}_{pz} \rightarrow 0$, or equivalently as $D_H \rightarrow 0$. If $D_P=0$, modes with $k=k_+$ have a longer wavelength relative to modes with 
$k=k_-$. The modes corresponding to $k=k_\pm$ represent damped ion cyclotron and whistler waves.
The $\pm$ signs outside the bracket correspond to propagation in either of two directions. 

At the photosphere the FAL background values give $\rho_0 \sim 2.74 \times 10^{-7}$ g-cm$^{-3}$, ${\bar{\nu}}_{pz} 
\equiv {\bar{\omega}}_{pz}/(2\pi) \sim 142.2$ Hz, $V_{A0} \sim 1.08$ km-sec$^{-1}$, and $D_P \sim 1.37 \times 10^8$ 
cm$^2$-sec$^{-1}$ for $B_z = 200$ G. Then $\omega/{\bar{\omega}}_{pz} \sim \nu/142 \; \mbox{Hz}$, and $\omega 
D_P/V_{A0}^2 \sim \nu/13 \; \mbox{Hz}$. Here ${\bar{\omega}}_{pz} \propto B_z$, and $D_P(0)/V_{A0}^2 \propto 
B_z^{-2}$ since $\Gamma(0) < 1$. The numerical solutions presented in this paper have $B_z \geq 500$ G and $\nu \leq 
3.4$ Hz. For this parameter range, $V_{A0} k_\pm/\omega \sim 1 \mp \omega/2{\bar{\omega}}_{pz}- i \omega D_P/2V_{A0}^2$,
where $\omega/2{\bar{\omega}}_{pz}$ and $\omega D_P/2V_{A0}^2$ are $\ll 1$. Then the condition $Re(k)<0$ implies that $k = - k_\pm$. 

Frequencies greater than $3.4$ Hz are not considered due to limited numerical resolution. Accurate computation of the solution at this frequency requires a numerical resolution $\lesssim 3.3$ m. The required numerical resolution
is determined mainly by the dissipative length scale $(D_P/\omega)^{1/2}$. This is discussed in more detail in 
\S 6.1.

\subsection*{3.2. $B_{y1\omega}(0)$ as a Function of $B_{x1\omega}(0)$}

Combining equations (31) and (34) in KG06 gives an equation for $B_{y1\omega}(0)$ as a function of $B_{x1\omega}(0),
\omega, k$, and the background state at $z=0$. Using $k(\omega)$ from equation (\ref{21}) to eliminate $k$ gives
\begin{equation}
B_{y1\omega}(0) = \mp \frac{i \rho(0)}{m_p n(0)} B_{x1 \omega}(0) \sim \mp i B_{x1 \omega}(0). \label{22}
\end{equation}
The factor $\rho(0)/m_p n(0) \sim 1.28$. It is set equal to unity, with small error, to be consistent with the equations for the un-coupled $k_+$ and $k_-$ modes presented in \S 5. As discussed in more detail in \S 5, since $k_+$ and $k_-$ are
almost identical for the parameter ranges of $B_z$ and $\nu$ considered here, only one set of modes needs to be considered.
This significantly simplifies solving equations (\ref{6}) and (\ref{7}).

The $\mp$ signs in equation (\ref{22}) correspond to the $\pm$ signs inside the bracket in equation (\ref{21}) for $k$. The
coupling of $B_{y 1 \omega}$ and $B_{x 1 \omega}$ is due to the Hall conductivity $\sigma_H$. If $\sigma_H$ is set equal
to zero, equation (\ref{22}) does not exist, in which case $B_{x1\omega}(0)$ and $B_{y1 \omega}(0)$ become un-coupled and
independent of one another. This is seen in general from equations (10) and (11) in KG06, which are the Faraday law equations for $B_x$ and $B_y$. In the linear approximation, each of those equations is coupled to the other one through the term 
$\propto (c^2/4\pi) \sigma_H/(\sigma_P^2+\sigma_H^2) = - c^2 \eta_H/4 \pi = D_H$. 

\subsection*{3.3. $B_{x1\omega}^\prime(0)$ and $B_{y1\omega}^\prime(0)$ as Functions of $B_{x1\omega}(0)$}

Equations (\ref{4}) and (\ref{5}) give $B_{x1 \omega}^\prime(0)$ and $B_{y1 \omega}^\prime(0)$ in terms of 
$V_{x1 \omega}(0)$ and $V_{y1 \omega}(0)$. Use equations (30) and (31) of KG06 to obtain $V_{x1 \omega}(0)$ and 
$V_{y1 \omega}(0)$ in terms of $B_{x1 \omega}(0)$ and $B_{y1 \omega}(0)$. Then use equation (\ref{22}) for 
$B_{y1\omega}(0)$ in terms of $B_{x1 \omega}(0)$. This gives
\begin{eqnarray}
B_{x1 \omega}^\prime(0) &=& i k B_{x1 \omega}(0) \label{23} \\
B_{y1 \omega}^\prime(0) &=& \pm k B_{x1 \omega}(0).  \label{24}
\end{eqnarray}
Here the $\pm$ signs in equation (\ref{24}) correspond to the $\mp$ signs in equation (\ref{22}). Then modes with $k=\pm k_+$ have 
$B_{y1 \omega}(0)=-i B_{x1 \omega}(0), B_{x1 \omega}^\prime(0)=\pm i k_+ B_{x1 \omega}(0),B_{y1 \omega}^\prime(0) = 
\pm k_+ B_{x1 \omega}(0)$, and modes with $k=\pm k_-$ have $B_{y1 \omega}(0)=i B_{x1 \omega}(0), B_{x1 \omega}^\prime(0)=\pm i k_- B_{x1 \omega}(0),B_{y1 \omega}^\prime(0) = \mp k_- B_{x1 \omega}(0)$. These two sets of boundary conditions are consistent with equation (\ref{22}), and show that the $\mp$ signs in that equation correspond to $k=k_\pm$.  

The solution for $B_{x1 \omega}(z)$ and $B_{y1 \omega}(z)$ is then determined as follows. Specify $B_{x1 \omega}(0)$. Use
equations (\ref{22})-(\ref{24}) to determine $B_{x1 \omega}^\prime(0), B_{y1 \omega}(0)$, and $B_{y1 \omega}^\prime(0)$.
Specifying $B_{x1 \omega}(0)$ determines the solution.

Without loss of generality choose $B_{x1}(0,t)$ to oscillate as $\cos \omega t$. Since $B_{x1}(0,t)= Re(B_{x1\omega}(0))
\cos \omega t - Im(B_{x1\omega}(0)) \sin \omega t$ it follows that $Im(B_{x1\omega}(0))=0$, so $B_{x1 \omega}(0)$ is real.

\section*{4. Resistive Heating Rate}

The heating rate per unit volume may be written as $Q= Q_\parallel + Q_\perp$. Here $Q_\parallel$ and $Q_\perp$ are the
heating rates due to dissipation of magnetic field aligned currents, and currents $\perp {\bf B}$, respectively. The
latter heating rate is due to Pedersen current dissipation. The Pedersen current is the current parallel to ${\bf E}_{CM \perp}$.
Previous modeling indicates that $Q_\parallel$ is orders of magnitude less than $Q_\perp$ in the chromosphere (KG06,
Goodman 2000, 2001, 2004 a,b). Exact expressions for $Q_\parallel$ and $Q_\perp$ are given by equations (23) and (24) of
KG06. In the linear approximation these equations show that $Q_\parallel$ is fourth order in the perturbation. The lowest order terms in $Q_\perp$ are second order. Then through second order in the perturbation the average of $Q$ over a period is 
\begin{eqnarray}
\langle Q(z,t) \rangle &=& \left(\frac{c}{4\pi}\right)^2 \frac{(1+\Gamma_0(z))\eta_{\parallel 0}(z)}{2} \left(B_{x1\omega}^\prime(z) \cdot B_{x1\omega}^\prime(z) + \right. \nonumber \\
& & \left. B_{y1\omega}^\prime(z) \cdot B_{y1\omega}^\prime(z) \right) \label{25} \\
\lefteqn{ =\frac{(1+\Gamma_0)\eta_{\parallel 0}}{2} \left(J_{x1\omega}(z) \cdot J_{x1\omega}(z) + J_{y1\omega}(z) \cdot J_{y1\omega}(z) \right)}  \label{26}.
\end{eqnarray}
This equation follows from the expression for the current density ${\bf J}_{1\omega} = (c/4\pi)(- B_{y1\omega}^\prime \hat{\bf x}+ B_{x1\omega}^\prime \hat{\bf y})$.

Combined with the results in \S 3.3, equation (\ref{25}) implies that $\langle Q(z,t) \rangle \propto 
B_{x1 \omega}^2(0)$ for fixed $B_z$ and $\nu$. 

\section*{5. Equations for Uncoupled $k_\pm$ Modes}

Inspection of equations (\ref{6}) and (\ref{7}) shows that the assumption $B_{y 1 \omega}(z)=\pm i B_{x1 \omega}(z)$ makes those equations identical to the single equation 
\begin{equation}
\left(\left(D_P(z) - \frac{i V_A^2}{\omega} \mp i D_H(z)\right) 
B_{x1 \omega}^\prime(z)\right)^\prime =  i \omega B_{x1\omega}(z) \label{27}
\end{equation}

The meaning of this simplified form of equations (\ref{6}) and (\ref{7}) follows from the boundary conditions chosen in 
\S 3. It follows that equation (\ref{27}) and the equation $B_{y 1 \omega}(z)=\pm i B_{x1 \omega}(z)$ describe the two
separate modes having $k=\pm k_+$ and $k=\pm k_-$. Solutions with $B_{y 1 \omega}(z)= i B_{x1 \omega}(z)$ correspond to 
$k=\pm k_-$. Solutions with $B_{y 1 \omega}(z)= -i B_{x1 \omega}(z)$ correspond to $k=\pm k_+$.

Since $k_+$ and $k_-$ are almost identical for the values of $B_z$ and $\nu$ considered here, the properties of these two modes are almost identical, so it is only necessary to determine the properties of one mode, chosen here to be the $k_+$ mode. Requiring that $S_z(0)>0$, which implies $S_z(z)>0$ for $z \geq 0$, requires choosing $k=-k_+$. Equation (\ref{27})
with the $+$ sign in front of $D_H$ determines the solutions presented in this paper.

Frequencies for which $k_+$ and $k_-$ are significantly different, corresponding to a large splitting of the 
Alfv\'{e}nic oscillations into whistler and ion cyclotron oscillations, and cases in which $S_z(z)<0$, corresponding to an electromagnetic energy flux downward from the corona might be important, but are not considered here.

\section*{6. Numerical Solution of the Model}

Equation (\ref{27}) is solved as follows. Let
\begin{equation}
A_\pm(z)= -\left(\frac{V_A(z)}{\omega}\right)^2 \pm \frac{D_H(z)}{\omega} - i \frac{D_P(z)}{\omega}. \label{28}
\end{equation}
Here $\pm$ corresponds to $k_\pm$, and it is noted that $D_H(z) < 0$. Integrating equation (\ref{27}) once gives
\begin{equation}
B_{x1 \omega}^\prime(z)=\frac{1}{A_\pm(z)}\left(A_\pm(0) B_{x1 \omega}^\prime(0) + \int_0^z B_{x1 \omega}(\alpha) \;
 d\alpha \right) \label{29}
\end{equation}

For a sufficiently small height increment $\Delta z$,
\begin{equation}
B_{x1 \omega}(z+\Delta z) \sim B_{x1 \omega}(z) + B_{x1 \omega}^\prime(z) \Delta z. \label{30}
\end{equation}
Here $B_{x1 \omega}^\prime(z)$ is given by equation (\ref{29}) with the integral approximated by a discrete sum with
$d \alpha \rightarrow \Delta z$. Given $B_{x1 \omega}(0)$, and using equation (\ref{23}), equations (\ref{29}) and 
(\ref{30}) may be solved iteratively to determine $B_{x1 \omega}(z)$ and $B_{x 1 \omega}^\prime(z)$. It remains to 
determine how small $\Delta z$ must be to ensure an accurate solution. This depends on the intrinsic length scales of the model.

\subsection*{6.1. Intrinsic Length Scales, and Required Numerical Resolution}

The meaning of a numerical solution to an MHD model can only be determined in the context of knowing the numerical spatial resolution, and the effects of any ad hoc diffusion terms and parameter values used in the computer code to increase stability, and allow for a solution to be obtained within a reasonable period of time. These artificial quantities, and the nonzero resolution, effectively change the model equations being solved, and introduce spurious effects into the solution. These effects can be large. They are related to the issue of including the relevant transport coefficients in the model, resolving their effects with sufficient accuracy, and being able to distinguish them from spurious effects, ideally by minimizing the spurious effects.

\subsubsection*{6.1.1. Intrinsic Length Scales}
Equation (\ref{27}) is simple enough so it can be solved numerically using a spatial resolution sufficiently high to
generate an accurate solution. Divide equation (\ref{27}) by $i \omega$, and decompose it into
real and imaginary parts. The second derivative terms in the two resulting real equations are multiplied by the squares
of the lengths $L_P \equiv (D_P/\omega)^{1/2}$ and $L_H \equiv (V_A/\omega) (1+\omega|D_H|/V_A^2)^{1/2}$. These are the
intrinsic length scales of the model. $L_P$ is a resistive length scale. $L_H$ equals the ideal MHD length scale 
$V_A/\omega$, which is the ideal MHD wavelength, times a dispersive correction factor that involves the square of the ratio of the dispersive length scale $(|D_H|/\omega)^{1/2}$ to the ideal MHD length scale. 

$L_P/L_H$ obeys the following scaling relations with respect to $B_z$ and $\nu$: (1) If $1+ \omega |D_H|/V_A^2 \sim 1$ then (a) $L_P/L_H \propto \nu^{1/2}$ if $1+ \Gamma \sim \Gamma$, and (b) $L_P/L_H \propto \nu^{1/2}/B_z$ if $1+ \Gamma \sim 1$. The condition $1+ \Gamma \sim \Gamma$ holds in the weakly ionized, strongly magnetized chromosphere. The condition $1+ \Gamma \sim 1$ holds in the weakly ionized, weakly magnetized photosphere, and in the strongly ionized, strongly magnetized corona. (2) If $1+ \omega |D_H|/V_A^2 \sim \omega |D_H|/V_A^2$ then (a) $L_P/L_H \propto B_z^{1/2}$ if $1+ \Gamma \sim \Gamma$, and (b) $L_P/L_H \propto 1/B_z^{1/2}$ if $1+ \Gamma \sim 1$.\footnote{$L_P/L_H$ also depends 
on the temperature and particle densities of the FAL background state, shown in figure 1. Here the focus is on the variation of $L_P/L_H$ with $\nu$ and $B_z$.}

These scaling relations may be combined with figure 2 to determine the value of $L_P/L_H$ as a function of $B_z,\nu$,
and $z$. Figure 2 shows $L_P$ and $L_H$ for $B_z=10^3$ G, and $\nu=100$ mHz. For the ranges of $B_z$ and $\nu$ considered in this paper, $L_P \ll L_H$, so $L_P$ defines the smallest scale that must be resolved. 

For the case of figure 2, the Hall dispersion term $\omega |D_H|/V_A^2$ increases from $\sim 10^{-4}$ at $z=0$ to 
$\sim 10^{-3}$ at $z = 850$ km, and then decreases rapidly to $\sim 10^{-7}$ at $z=2000$ km. Then for $B_z=1000$ G, 
$\nu$ must increase to $\sim 10^3-10^4$ Hz before Hall dispersion becomes significant. However, as indicated in \S 2, the
condition $\nu \ll 720$ Hz must be satisfied in order that the Ohm's law used here be valid. If $\nu \gtrsim 720$ Hz,
either a more general form of the Ohm's law may be used, or, at sufficiently high frequencies, a kinetic or multi-fluid
model is necessary to accurately determine the resistive heating rate. Then the Ohm's law used here is valid for 
$\nu \lesssim 10^2$ Hz. If $B_z$ is reduced to 100 G, Hall dispersion becomes important for $\nu \sim 10^2-10^3$ Hz,
which is within the frequency range for which the Ohm's law is valid.

\subsubsection*{6.1.2. Required Numerical Resolution}

For given values of $B_z$ and $\nu$ it is necessary to use a numerical resolution $\Delta z \ll L_P(z)$ at a given height to accurately compute $Q(z)$. The reason is that each component of ${\bf J}$ is computed by taking a
difference of derivatives of components of ${\bf B}$. This doubly compounds the numerical error present in ${\bf B}$. Since 
$Q \propto J^2$, additional error is introduced by compounding the error in ${\bf J}$ by taking its square.\footnote{A related problem is estimating ${\bf J}$ from observations of ${\bf B}$. The finite resolution of the observations causes an error in the estimate of ${\bf J}$. The meaning of this estimate can be determined only in the context of knowing the error in the estimate. Knowing this error is especially important when correlating observed emission with observational estimates of ${\bf J}$. For example, the vector magnetic field observations used by Socas-Navarro (2005a,b) to correlate current density with chromospheric emission in a sunspot have a spatial resolution $\sim 0.6^{''}$, in which case only current densities with spatial scales $\gtrsim 4 \times 0.6^{''} \sim 1700$ km can be detected. If the heating that causes emission primarily occurs on smaller scales, there is little or no observed correlation between emission and current density, which is the result obtained by Socas-Navarro (2005 b).} In addition,
the solution at a height $z_0$ depends on the solution at heights $0 \leq z < z_0$ since the solution is generated by
integrating upward from the photosphere. Then $\Delta z(z)$ needs to be chosen $\ll L_P(z)$ over any given height range in order to generate an accurate solution in the overlying atmosphere. 

Figure 2 shows that $0.2 \lesssim L_P(z) \lesssim 1$ km for $0 \leq z \leq 500$ km, $1 \lesssim L_P(z) \lesssim 10$ km
for $500 \leq z \leq 1000$ km, and $10 \lesssim L_P(z) \lesssim 100$ km for $1000 \leq z \leq 2000$ km. Most
chromospheric heating occurs below $z=1000$ km, and, as shown in the numerical examples in \S 6.3, a large fraction of
the Poynting flux that flows through the photosphere into the overlying atmosphere can be thermalized by electron current
dissipation below $z \sim 500$ km. This dissipation can have a strong effect in regulating the Poynting flux
that reaches the chromosphere, which affects the chromospheric and coronal heating rates. Together with the variation of 
$L_P(z)$ with height in figure 2, this suggests that $\Delta z(z)$ must be chosen so that: (1) $\Delta z \ll 0.1$ km below $z=500$ km to accurately compute the heating rate due to electron current dissipation; (2) $\Delta z \ll 1$ km for $500 \leq z \leq 1000$ in order to accurately compute most of the chromospheric heating rate; (3) $\Delta z \ll 10$ km for $1000 \leq z \leq 2000$ in order to accurately compute the remainder of the chromospheric heating rate. Since $\Delta z$ is constant in the model used here, these estimates suggest that $\Delta z$ must be chosen $\ll 0.1$ km. This is found to be the case for the numerical examples presented in \S 6.3. For example, for each value of $B_z=(500,700,1000,2000,3000)$ G, as $\nu$ is respectively increased to $(0.5,0.8,1.3,3.0,3.4)$ Hz, it is found necessary to reduce $\Delta z$ to 
$\sim 3.33$ m in order to prevent the generation of spatial numerical oscillations. The choice $\Delta z =3.33$ m is made for all numerical examples presented in 
\S 6.3. 

The characteristic values of $L_P(z)$ for the model presented here, and the corresponding need to use relatively small
values of $\Delta z$ to compute an accurate solution are not restricted to the model considered here. Similar
values of $L_P(z)$ must appear in more general models, such as nonlinear, multi-dimensional models that simulate processes with characteristic time scales $\Delta t \sim 1/\nu$ corresponding to the frequencies used here, since
these values of $L_P(z)$ arise from characteristic values of the temperature, densities, and magnetic field strength in the photosphere and chromosphere. In addition, if $B$ decreases with height, 
$L_P(z)$ decreases in the chromosphere, where it is $\propto B$. Since numerical errors propagate and
may be amplified in time and space, the upper bounds on the general grid spacings $\Delta x(x,y,z,t), \Delta y(x.y.x,t)$, and $\Delta z(x,y,z,t)$ are expected to be smaller than those that apply to the model considered here. Given the required values of $\Delta z(z)$ estimated in the previous paragraph, this places severe constraints on the spatial resolution required in multi-dimensional models in order to compute meaningful resistive heating rates in the photosphere and chromosphere. 

Figure 2 also shows a general property of solutions to the model that is discussed quantitatively in \S 6.3. This property is as follows. In the region below the height of the temperature minimum,
which is the lower boundary of the chromosphere, resistive heating is mainly due to dissipation of electron currents. In the region above the temperature minimum resistive heating is mainly due to dissipation of
proton Pedersen currents. 

\subsection*{6.2. Poynting Flux}

Let $k_r$ and $k_i$ be the real and imaginary parts of $k$ in equation (\ref{21}). Let $S_z(z,t)$ be the vertical component of the Poynting flux ${\bf S}=c ({\bf E} \times {\bf B})/4 \pi$. The time average of $S_z$ at the photosphere is
\begin{equation}
\langle S_z(0,t) \rangle = -\frac{|B_{x1\omega}(0)|^2}{8 \pi} \left\{\frac{2 V_{A0}^2(0) k_r}{\omega} +
\frac{c^2 \eta_{\parallel 0}(0)}{2 \pi} \left(- k_i \left(1+\Gamma_0(0)\right) \pm k_r M_{e0}(0) \right) \right\}. 
\label{31}
\end{equation}
Here the $\pm$ signs correspond to $|k|= |k_\pm|$.
The term in equation (\ref{31}) that is $\propto \eta_{\parallel 0}$ is the resistive term. The other term is the ideal MHD term, modified by the effect of the Hall conductivity, which gives rise to the factor of 2 in this term. This term is positive since $k_r < 0$. The ratio of the resistive term to the ideal MHD term is $\sim D_P \omega/V_{A0}^2 \sim
(\nu/13 \; \mbox{Hz})(200 \; \mbox{G}/B_z)$, using the estimates in the second paragraph following equation (\ref{21}).
This ratio is $\ll 1$ for the ranges of $\nu$ and $B_z$ considered here. The aforementioned estimates also imply that
$k_r \sim - \omega/V_{A0}$. Then, $\langle S_z(0,t) \rangle \sim V_{A0}(0)|B_{x1\omega}(0)|^2/4 \pi > 0$. If $k_r$ is
chosen to be positive, then $\langle S_z(0,t) \rangle$ has the same magnitude, but is negative.

The MHD form of Poynting's theorem is 
\begin{equation}
\frac{\partial}{\partial t}\left(\frac{B^2}{8\pi}\right)+\nabla \cdot {\bf S} = - Q - {\bf V}\cdot {\bf F}_m. \label{32}
\end{equation}
Here ${\bf F}_m = ({\bf J} \times {\bf B})/c$ is the magnetic Lorentz force acting on the CM of a fluid element. Then ${\bf V}\cdot {\bf F}_m$ is the rate at which energy is exchanged between the electromagnetic and CM kinetic energy reservoirs. For the model considered here $\langle {\bf V}\cdot {\bf F}_m \rangle = 
\langle \partial B^2/\partial t \rangle =0$ through second order in the perturbation, using the fact that the vertical component of the perturbed magnetic field is zero. Then equation (\ref{32}) implies that
$\nabla \cdot \langle {\bf S} \rangle = - \langle Q \rangle$. Integrating this equation gives
\begin{equation}
\langle S_z(z,t)\rangle = \langle S_z(0,t)\rangle - \int_0^z \langle Q(\alpha,t)\rangle \; d\alpha. \label{33}
\end{equation}

For $B_{y1 \omega}(z)=-i B_{x1 \omega}(z)$, which is the case considered here, the period averaged vertical Lorentz force is $-(|B_{x1 \omega}(z)|^2)^\prime/8 \pi > 0$. A similar result is derived in DePontieu \& Haerendel (1998). It cannot be concluded from this result that the oscillations drive a net vertical mass flux $\langle \rho V_z \rangle$
because this model, like the one in DePontieu \& Haerendel (1998) does not include density or vertical velocity perturbations. The reason is that the absence of a horizontal component of the background magnetic field uncouples $V_z$ and $\rho$ from the magnetic oscillations. Within the context of a 1 D linear model, the background field must have a horizontal component in order to compute the net vertical mass flux due to magnetic oscillations. Adding such a component couples compressible MHD modes to the Alfv\'{e}nic modes, so the oscillations are no longer purely Alfv\'{e}nic.

Since $\langle {\bf V} \cdot {\bf F}_m \rangle =0$, it follows that $\langle Q \rangle \equiv \langle {\bf J} \cdot 
({\bf E} + ({\bf V} \times {\bf B})/c) \rangle =  \langle {\bf J} \cdot {\bf E} \rangle$. Then the heating is not convection driven, meaning there is no net flow of CM kinetic energy into thermal energy. Although ${\bf J} 
\cdot ({\bf V} \times {\bf B})/c = - {\bf V} \cdot {\bf F}_m \neq 0$, so there is a time dependent exchange of energy
between the CM kinetic energy reservoir and the current density, its average over a period
is zero.

\subsection*{6.3. Numerical Examples}

The height range of the solutions extends from $z=0$ to $z=19000$ km. For all examples the height of the base of the chromosphere is defined as the height $z_1$ where $\Gamma=1$, and hence where 
$\eta_P=2 \eta_\parallel$. This is the height at which the magnetization, and hence $\Gamma$ and $\eta_P$, begins to increase rapidly with height. This height varies with $B_z$. Larger (smaller) values
of $B_z$ correspond to smaller (larger) values of $z_1$, and larger (smaller) values of the chromospheric heating flux.
The top of the chromosphere is defined to be at $z= 2156.7$ km, which is just below the FAL TR. 

The period averaged Poynting fluxes through the photosphere, and into the upper corona are $S_{Ph} \equiv 
\langle S_z(0,t) \rangle$, and $S_{Cor}\equiv \langle S_z(19000 \;\mbox{km},t)\rangle$. The period averaged photospheric, chromospheric, and coronal heating fluxes $F_{Ph},F_{Ch}$, and $F_{Cor}$ are respectively defined as the integrals of 
$\langle Q(z,t) \rangle$ over the height ranges $0 \leq z \leq z_1,z_1 \leq z \leq 2156.7 \; \mbox{km}$, and $2156.7 \; \mbox{km} \leq z \leq 19000 \; \mbox{km}$. These fluxes satisfy $S_{Ph} = F_{Ph} + F_{Ch} + F_{Cor} +  S_{Cor}$.

For all examples the highest frequency used is such that using much higher frequencies causes significant numerical noise to appear for the fixed numerical resolution of 3.33 m.

\subsubsection*{6.3.1. Solution for $B_z=10^3$ G, $B_{x1 \omega}(0)= 0.1 B_z$}

Figures 3-11 describe the solution. For this solution the lower boundary of the chromosphere is at $z_1 = 349$ km. For comparison, $\Gamma$ reaches the value 5 at $z=438$ km.

Figure 3 shows the Poynting and heating fluxes as a function of frequency. The Poynting flux flowing upward into the atmosphere through the photospheric surface $z=0$ is $S_{Ph} = 4.3 \times 10^8$ ergs-cm$^{-2}$-sec$^{-1}$. It is essentially independent of $\nu$ for fixed $B_{x1 \omega}(0)$, as follows from the discussion immediately after equation 
(\ref{31}). The figure shows that as $\nu$ increases, $F_{Ph}$ and $F_{Ch}$, respectively due to resistive dissipation of electron currents and ion Pedersen currents, initially increase while $S_{Cor}$ decreases since increasingly more energy is being thermalized in the underlying atmosphere. $F_{Ch}$ reaches its maximum value of $5.6 \times 10^7$ ergs-cm$^{-2}$-sec$^{-1}$ at $\nu= 780$ mHz, and then decreases with increasing frequency as increasingly more Poynting flux is thermalized in the photosphere, indicated by the increase in $F_{Ph}$ towards $S_{Ph}$. Electron current dissipation in the photosphere may play an important role in regulating the Poynting flux into the overlying atmosphere, which in turn regulates $F_{Ch}$. The figure shows that 
$F_{Ph} \sim 3-13 F_{Ch}$, and suggests that as $\nu$ increases above $1300$ mHz, $F_{Ph} \rightarrow S_{Ph}$. In this limit essentially all of the Poynting flux injected into the
atmosphere is thermalized in the photosphere by electron current dissipation.

For figures 4-7, $\nu=780$ mHz, which is the frequency at which $F_{Ch}$ is a maximum.

Figure 4 shows the real and imaginary parts of $B_{x1 \omega}(z)$. The amplitudes decrease almost exponentially with height up to $z \sim 800$ km, with a scale height $\sim 327.4$ km. The amplitudes damp down to near zero at the top of the chromosphere, and continue to decrease into the corona. $B_{y 1 \omega}(z)$ has essentially the same spatial dependence as $B_{x1 \omega}(z)$. Most of the damping in the chromosphere occurs below $10^3$ km.

Figure 5 shows the real and imaginary parts of $V_{x1 \omega}(z)$, where $V_{x1 \omega R}(0)= 
-0.54$ km-sec$^{-1}$, and $V_{x 1 \omega i}(0)= 0$. The amplitudes grows with increasing height due to the decrease in density since $V_{x1 \omega} \propto B_{x1 \omega}^\prime/\rho_0$. The amplitude $V_{y 1 \omega}(z)$ behaves in a similar way. The RMS velocity perturbation amplitude, defined by $V_{rms}(z)= [(|V_{x1 \omega}(z)|^2 + 
|V_{y 1 \omega}(z)|^2)/2 ]^{1/2}$, increases to $\sim 140$ km-sec$^{-1}$ at $z=19000$ km.

Figure 6 shows the period averaged total and Spitzer heating rates, the Pedersen and Spitzer resistivities, and the RMS
current density. The Spitzer heating rate is defined by $Q_S=\eta_\parallel J^2$, so $Q=Q_S + \Gamma \eta_\parallel
J_\perp^2$. 

The total and Spitzer heating rates are equal, nearly constant, and have their maximum value up to a height $\sim 
z_1 = 349$ km where they begin to decrease, but $\langle Q \rangle$ decreases much more slowly than $\langle Q_S \rangle$ due to the rapidly increasing magnetization of the gas. $\langle Q \rangle$ is orders of magnitude greater than 
$\langle Q_S \rangle$ throughout the chromosphere. They become equal again in the TR and corona where the degree of ionization is so strong that $\Gamma < 1$ despite the plasma being strongly magnetized. The Pedersen and parallel resistivities behave in a similar manner. They are nearly equal and have their minimum value below the TR up to a height $\sim z_1$, and then diverge rapidly as $\eta_P$ increases due to increasing magnetization. The high magnetization of the chromosphere causes it to be a highly resistive gas with respect to the Pedersen current, which flows parallel to the driving electric field ${\bf E}_{CM \perp}$.

The large value of $\langle Q \rangle$ below $z_1$ is due to dissipation of electron currents. The corresponding heating flux is $\sim 200\; \mbox{km} \times 10 \; \mbox{ergs-cm$^{-3}$-sec$^{-1}$}= 2 \times 10^8$ 
ergs-cm$^{-2}$-sec$^{-1}$. This is $\sim 46 \%$ of the Poynting flux through the photosphere, shown in figure 3. Almost half of the Poynting flux is thermalized by electron current dissipation in the photosphere before it can reach the chromosphere. The current density is greatest in the photospheric region where electron current dissipation dominates the heating rate. Although $J_{rms}$ decreases by orders of magnitude with increasing height in the chromosphere, the large magnetization and correspondingly large Pedersen resistivity maintain the heating rate at a significant level.

Figure 7 shows properties of the electric field. Here $E_{x1\omega}=E_{x1\omega, conv}+E_{x1 \omega, res}$, which is the sum of convection and resistive components. Similarly for $E_{y1\omega}$, which is not shown. The figure shows that the convection component dominates the resistive component by two or more orders of magnitude, so $E_{x1\omega}$ and 
$E_{x1\omega, conv}$ are almost identical. In this sense the ideal MHD Ohm's law ${\bf E} + ({\bf V} \times 
{\bf B})/c =0$ is nearly valid. The departure from ideality is due to the presence of ${\bf E}_{res}$, which, though the magnitude of its components are small compared with those of ${\bf E}_{conv}$, is responsible for the resistive heating since 
$Q={\bf J} \cdot {\bf E}_{res}$. If ${\bf E}_{res}$ is small in this sense in more complex models that must be solved numerically with limited resolution, the question of whether ${\bf E}_{res}$, and hence the resistive heating rate, is accurately computed must be addressed.

Figure 8 shows the height variation of the RMS magnetic field perturbation amplitude for a range of $\nu$. The figure shows that the damping increases with $\nu$, and that almost all of the damping occurs below $10^3$ km as already indicated for the example in  figure 4. This is consistent with the profile of $\langle Q \rangle$ for $\nu=780$ mHz in figure 6. Figure 8 also shows that $B_{rms}^\prime$ rapidly becomes relatively small above $10^3$ km, consistent with the decrease in the RMS current density in figure 6. 

Figure 9 shows the height variation of the RMS velocity amplitude for the same range of $\nu$ used in figure 8. Overall, $V_{rms}$ initially increases, and then decreases with increasing 
$\nu$, and for the highest frequencies it eventually decreases with increasing height. This behavior is due to the interplay of three factors, and may be understood as follows. Recall that $V_{x1 \omega}(z) \propto
B_{x1 \omega}^\prime(z)/\nu \rho_0(z), V_{y1 \omega}(z) \propto B_{y1 \omega}^\prime(z)/\nu \rho_0(z)$, and $Q \propto \eta_P (B_{x1\omega}^\prime \cdot B_{x1\omega}^\prime + B_{y1\omega}^\prime \cdot 
B_{y1\omega}^\prime)$. The decrease in $\rho_0$ with height tends to increase $V_{rms}$, and the factor of $\nu^{-1}$ scales it down as $\nu$ increases. As $\nu$ initially increases, $F_{Ch}$ and $F_{Cor}$ increase, reach a maximum near the same frequency and then decrease. It follows from the expression for $Q$ that the magnitudes of $B_{x1 \omega}^\prime(z)$ and $B_{x1 \omega}^\prime(z)$, and hence of the current density, eventually decrease with increasing $\nu$, tending to reduce $V_{rms}$ along with the factor $\nu^{-1}$. Since $\rho_0$ decreases much more slowly with height in the corona than in the underlying atmosphere, it is less effective in amplifying $V_{rms}$ in the corona than in the underlying atmosphere.

Figures 8 and 9 show the perturbation is linear for the solutions considered in the sense that $B_{rms}/B_z \ll 1$, and
$V_{rms}/V_A \ll 1$, where $V_A$ is computed using the HI and He densities in figure 1, and the fact that $B_z$ is constant. The Alfv\'{e}n speed $V_A= 191 B(10^3\;\mbox{G}) n^{-1/2}(10^{14}\;\mbox{cm}^{-3})$ km-sec$^{-1}$, where $n = n_{HI} + n_{He} \sim 1.1 n_{HI}$, and an average mass of $1.3 m_p$ is used. 
The inequality $V_{rms}/V_A \ll 1$ follows from $B_{rms}/B_z \ll 1$ for the following reason. 
Equation (\ref{4}) may be written as $|V_{x 1 \omega}(z)|/V_A(z)= V_A(z) (|k(z)|/\omega)(|B_{x 1 \omega}(z)|/B_z)$ 
for some real function $k(z)$. Here $\omega/|k(z)|$ is expected to be on the order of $V_A(z)$, in which case $|V_{x 1 \omega}(z)|/V_A(z) \sim |B_{x 1 \omega}(z)|/B_z$. Then $|V_{x 1 \omega}|/V_A \ll 1$ if $|B_{x 1 \omega}(z)|/B_z \ll 1$. A similar analysis applies to equation (\ref{5}).

Figure 10 shows the height variation of $\langle Q \rangle$ for the same range of $\nu$. The figure shows that 
it increases with $\nu$ for $z \lesssim 10^2$ km, has its first and largest maximum in this region, and that it increases with $\nu$ for $\nu \leq 10^3$ mHz and $100 \lesssim z \lesssim z_1$. It is in the region $z \lesssim z_1$, 
especially the sub-region $z \lesssim 10^2$ km, where heating by electron current dissipation is strongest, and where it can have a strong effect in regulating the Poynting flux into the upper atmosphere. For $z > z_1$, $\langle Q \rangle$ tends to increase with $\nu$ up to $\nu \sim 700$ mHz, consistent with $F_{Ch}$ reaching its maximum at $\nu = 780$ mHz, and then decrease with increasing $\nu$ as a greater fraction of $S_{Ph}$ is thermalized by electron current dissipation in the photosphere. There is a second peak near $z=700$ km, also seen in figure 6. The presence of these peaks is due to the variation of $\eta_P$ with height. This is seen as follows. $\langle Q \rangle = \eta_P J_{rms}^2 =
(1+ \Gamma) \eta_\parallel J_{rms}^2$, and $dJ_{rms}/dz <0$. Then a peak in 
$\langle Q \rangle$ must be due to a local and sufficiently rapid increase in $\eta_P$. Near $z=0$, $\eta_P \sim \eta_\parallel \propto n_H/n_e$, and the first maximum in $\langle Q \rangle$ is due to the fact that for
FAL, $d(n_H/n_e)/dz >0$ in this region (e.g. see figure 3 of Goodman 2004a). The second peak occurs where $\eta_P \sim \Gamma \eta_\parallel$, and is due to the rapid increase in $\Gamma$ in the lower
chromosphere.

Figure 11 shows the height variation of the time averaged heating rate per unit mass $\langle Q_m \rangle \equiv \langle Q \rangle/\rho_0$ for the same range of $\nu$. The figure shows that for $400 \leq \nu \leq 1300$ mHz, $\langle Q_m \rangle \sim 10^9 - 10^{10}$ 
ergs-g$^{-1}$-sec$^{-1}$ in the height range of 1000-2000 km, consistent with the predictions of the Anderson \& Athay (1989) semi-empirical model. 

\subsubsection*{6.3.2. Variation of Heating Rates and Fluxes with $B_z$}

For $B_z < 700$ G, it is found that values of $F_{Ch} \geq 10^7$ ergs-cm$^{-2}$-sec$^{-1}$ cannot be achieved unless perturbation amplitudes $B_{x1\omega}(0) > 0.1 B_z$ are used. All Poynting and heating fluxes, and heating rates per unit volume and mass are $\propto B^2_{x1 \omega}(0)$.

Figures 12-19 show the Poynting and heating fluxes, and $\langle Q_m \rangle$ for $B_z=(500,750,2000,3000)$ G, where the corresponding $B_{x1\omega}(0)=(0.1,0.1,0.01,0.01) B_z$. 

Combined with the figures for the $B_z=1000$ G case, the figures show that the fluxes and heating rate increase with 
$B_z$, that the lower boundary $z_1$ of the chromosphere, defined as the height at which $\Gamma=1$, decreases with increasing $B_z$, and that the frequency at which $F_{Ch}$ is a maximum increases with $B_z$. For $B_z= (500,700,1000,2000,3000)$ G the maximum values of $F_{Ch}=(2 \times 10^6,1.1 \times 10^7,5.6 \times 10^7,1.2 \times 10^7,5.9 \times 10^7)$ ergs-cm$^{-2}$-sec$^{-1}$, occurring at $\nu=(305,475,780,1800,2720)$ mHz. The corresponding $z_1= (426,389,349,268,219)$ km. Then increasing $B_z$ from 500 to 3000 G lowers the base of the chromosphere by
about 207 km. Strong magnetization, heating by proton Pedersen current dissipation, and corresponding chromospheric emission should begin lower in the atmosphere in regions with higher field strength.

The coronal Poynting fluxes shown in figures 3, 12, 14, 16, and 18 have a range $\sim 3.5 \times 10^7 - 4 \times 10^8$
ergs-cm$^{-2}$-sec$^{-1}$. This is sufficient to provide the required energy input of $\sim 10^7$ ergs-cm$^{-2}$-sec$^{-1}$ to the corona in active regions (Withbroe \& Noyes 1977; Jordan 1981).

\section*{7. A Wave Generation Mechanism}

A question is how Alfv\'{e}n waves are generated at frequencies $\sim 1$ Hz in the photosphere. A mechanism based on
the perturbation of photospheric magnetic flux tubes with diameters $\sim 10$ km by resistive magnetic reconnection in current sheets with thicknesses $\sim 100$ m is proposed. The analysis in this section supports this proposition. 

\subsection*{7.1. Alfv\'{e}n Modes of a Solenoidal Flux Tube}

Assume an ideal MHD model modified by including the Spitzer resistive term in the Ohm's law,
which is ${\bf E} + ({\bf V} \times {\bf B})/c = \eta_\parallel {\bf J}$. The Hall term is omitted since, as shown in 
\S 3.1, it has a small influence on the Alfv\'{e}n mode spectrum in the photosphere. The resistive heating rate driven by these modes is due to electron current dissipation. It is computed in \S 7.4, and shown to be a significant fraction of the upward Poynting flux driven by these modes, which is computed in \S 7.3. However, as suggested by the analysis in \S 3.1, and as shown in more detail in this section, resistivity has a small effect on the real part of the mode frequencies, and resistive damping of the modes is small for time intervals $\lesssim \nu^{-1}$, for the frequencies of interest here. Therefore, resistivity has a small effect on the mode frequencies, but the modes drive a significant resistive heating rate.

Use cylindrical coordinates $(R,\theta,z)$, with $z$ height above the photosphere.
Consider a steady state cylindrical flux tube of radius $R_0$ with a constant vertical magnetic field with magnitude $B$
confined by a pressure difference satisfying the radial force balance jump condition $p_i+B^2/8 \pi = p_o$ across the
surface $R=R_0$. Here $p_i$ and $p_o$ are the pressures immediately inside and outside of the surface $R=R_0$, and the magnetic field immediately outside this surface is assumed to be zero, so there is an azimuthal surface current density $K_\theta = 
c B/4\pi$ at $R=R_0$. Inside the flux tube, the density, pressure and temperature are assumed independent of $R$, and the velocity is assumed to be zero. Except in \S 7.4, all calculations are assumed to be done for heights $z$ such that 
$0 \leq z \ll  L$, and for wavelengths $\lambda = V_A/\nu \ll L$, where $L \sim 150$ km is the pressure scale height at the photosphere. Estimates of the height dependence of certain quantities are used in \S 7.4 to estimate the
resistive heating flux in the photosphere driven by these modes.

Introduce a linear perturbation $f_1(R) \exp(i (\omega t - k z))$ for each physical quantity $f$, assuming constant temperature, and the ideal gas equation of state. The corresponding dispersion relation has an Alfv\'{e}nic branch and a magnetoacoustic branch. Only the Alfv\'{e}nic branch is considered here. The solution for the perturbation inside the flux tube that is finite at the origin is given by $\omega^2 - k^2 V_A^2 = i \omega \eta_\parallel c^2 k^2/(2 \pi), B_{z1}=B_{R1}=V_{z1}=V_{R1}=\rho_1 = p_1=0$, $V_{\theta 1}=- V_A B_{\theta 1}/B$, and $B_{\theta 1}(R)= b_{\theta 1} J_1(kR)$, where $J_1$ is the Bessel function of the first kind, and $b_{\theta 1}$ is a constant.

The solution to the dispersion relation is 
\begin{eqnarray}
\omega &=& \pm k V_A \left(1-\left(\frac{\eta_\parallel c^2 k}{4 \pi V_A}\right)^2\right)^{1/2} + i 
\frac{\eta_\parallel c^2 k^2}{4 \pi} \label{3000} \\
& \equiv& \pm \omega_r + i \omega_i. 
\end{eqnarray}
It is now shown that $|\omega_r| \sim k V_A$ and $\omega_i/\nu \ll 1$ for $\nu \lesssim 1$ Hz. This is done by setting $k= \omega/V_A$ on the right hand of equation (\ref{3000}), assuming $\omega$ is real in this expression, and then showing
that the resistive terms are small. At the photosphere $\eta_\parallel \sim 1.91 \times 10^{-12}$ sec, and $V_A \sim 6.9$ km-sec$^{-1}$ assuming $B=10^3$ G. First, $\omega_i/\nu \sim 0.012 \nu$, so for $\nu \lesssim 1$ Hz, $\gtrsim 83$ waves are generated before resistive damping becomes significant. Next, $(\eta_\parallel c^2 k/(4 \pi V_A))^2 \sim 3.26 \times 
10^{-6} \nu^2$, so $|\omega_r| \sim k V_A$ with high accuracy. Henceforth, it is assumed that $\omega = V_A k$.      
Resistive effects are not considered again until \S 7.4. 

Now assume the perturbation is radially localized within the flux tube in the sense that for $R > R_0$, $|B_{\theta 1}(k R)|$ decreases to values $\ll$ its maximum value for $R < R_0$, and does so over a distance $\ll R_0$. This type of perturbation might occur when the flux tube is perturbed by an anti-parallel magnetic field $- B_o \; \hat{\bf z}$, with $B_o >0$, that is convected to the boundary of the flux tube, forming a current sheet centered near the surface $R=R_0$.
The current sheet undergoes magnetic reconnection that excites normal modes in the flux tube. As a result of this process
the surface current $K_\theta$ increases to $c (B+B_o)/4 \pi$. This type of normal mode excitation mechanism is 
discussed in more detail in \S 7.2.

Consider the limiting case for which the magnetic field perturbation is completely localized inside the flux tube. Then
the boundary condition $B_{\theta 1}(k R_0)=0$ must be satisfied, so $J_1(k R_0)=0$. Then $k R_0 = (3.832, 7.016, 10.173,...)$, which are the zeros of $J_1$. Since $k= 2 \pi/\lambda$, the boundary condition couples $\lambda$, and hence $\nu$, to $R_0$. At the photosphere $V_A \sim 6.9$ km-sec$^{-1}$ for $B=10^3$ G. Then the dispersion relation combined with the boundary condition implies $R_0(\mbox{km}) \sim (4.2, 7.7, 11.2,...)/\nu \equiv S/\nu$, where $S \equiv(4.2, 7.7, 11.2,...)$. For $\nu \sim 1$ Hz, the corresponding values of $R_0$ are consistent with the observations cited in \S 1 that set an upper bound $\sim 32.4$ km on the radii of magnetic flux tubes associated with G-band bright points.

The conclusion of the section is that the frequencies of the Alfv\'{e}nic oscillations considered earlier in this paper as drivers of chromospheric heating are consistent with those of the Alfv\'{e}nic normal modes of small scale magnetic flux tubes in the photosphere. 

\subsection*{7.2. Excitation of the Alfv\'{e}n Waves by Magnetic Reconnection}

Assume the flux tube is perturbed by magnetic reconnection occurring in current sheets formed at the boundary of the flux tube. A possible configuration is that of reconnection in a current sheet centered near $R=R_0$, and parallel to
the $z$ axis. Then the thickness of the sheet is along the radial direction. The flux tube field near the boundary serves as one of the anti-parallel field components that reconnect. Let the characteristic length over which the reconnecting field varies be $L_B$. This is the thickness of the current sheet. Assume the reconnection occurs via the resistive tearing mode. Then $L_B$ is estimated as follows, with the result that the current is thin in the sense that 
$L_B \ll 2 R_0$. 

The characteristic growth time $\tau$ for this mode is
given by $\tau=L_B^2 (k^2/(D^3 V_A^2))^{1/5}$ (e.g. Parker 1994). Here $2 \pi/k$ is the wavelength of the mode parallel to the current sheet, and the diffusivity $D= \eta_\parallel c^2/4 \pi$.
At the photosphere $D \sim 1.37 \times 10^8$ cm$^{2}$-sec$^{-1}$. Setting $\tau = 1/\nu, k=2 \pi \nu/ V_A,d=2 R_0$, and
using the boundary condition $R_0(\mbox{km})=S/\nu$ derived in \S 7.1, the expression for $\tau$ gives
\begin{equation}
\frac{L_B}{d} \sim \frac{\nu^{3/10}}{5 S} = \nu^{3/10} (4.8 \times 10^{-2}, 2.6 \times 10^{-2}, 1.8 \times 10^{-2},...).  \label{2000}
\end{equation}
Then for $\nu \sim 1$ Hz, $L_B/d \sim 10^{-2}$ for the first few modes, and is smaller for higher order modes.\footnote{The equation $R_0=S/\nu$ derived in \S 7.1 shows that lower frequency waves with $\nu < 1$ Hz, which experience less resistive dissipation in the photosphere and chromosphere, correspond to larger diameter flux tubes. Equation (\ref{2000}) shows that $L_B/d \ll 1$ for these waves.
Lower frequency waves carry a larger fraction of their initial energy into the corona.}

The conclusion of this section is that magnetic reconnection by the resistive tearing mode in a relatively thin current
sheet at the boundary of the flux tube is a possible excitation mechanism of the Alfv\'{e}n waves.

\subsection*{7.3. Poynting Flux of the Alfv\'{e}n Waves}

In this section it is shown that the upward Poynting flux of the Alfv\'{e}n waves in the flux tubes is consistent with the values of $S_{Ph}$ in figure 3 if the flux tubes have a filling factor $\sim 4 \%$, and $S_{Ph}$ is
assumed to be an average over areas with diameters $\sim 100 - 400$ km. 

The Poynting flux only has a vertical component. Its average over a period is $\langle S_z \rangle(kR) = V_A |b_{\theta 1}|^2 J_1^2(kR)/8 \pi$. Averaging this over a flux tube
area $\pi R_0^2$, with $kR_0$ given by the sequence of zeros of $J_1$ gives $\langle S_z \rangle_{avg}= |b_{\theta 1}(10^2 \; \mbox{G})|^2 (4.45 \times 10^7, 2.47 \times 10^7, 1.71 \times 10^7, ...)$ ergs-cm$^{-2}$-sec$^{-1}$. 
Then for $|b_{\theta 1}| \sim 100$ G $(= 0.1 B)$, $\langle S_z \rangle_{avg}$ is roughly 10 times smaller than $S_{Ph}=4.3 \times 10^8$ ergs-cm$^{-2}$-sec$^{-1}$ in figure 3.\footnote{The maximum value of $J_1(kR)$ is $0.6$ for $R \leq 
R_0$. Then the maximum relative amplitude of the magnetic field perturbation is $\sim 0.6 |b_{\theta 1}|/B = 0.06$ for
$|b_{\theta 1}|=100$ G, consistent with the assumption of a small amplitude perturbation.}

Now require that for a given filling factor $f$, the total Poynting flux from some area $A \equiv \pi R_\ast^2$
equals a given flux $F$. Then $F= R_\ast^2 f \langle S_z\rangle_{avg}/R_0^2$, so that $R_\ast = R_0 \left(F/f \langle S_z
\rangle_{avg} \right)^{1/2}$. Choose $F= 10 \langle S_z \rangle_{avg}$, so it is comparable to $S_{Ph}$ in figure 3, and choose 
$f=0.04$ consistent with observations cited in \S 1. Then $2 R_\ast = 31.62 R_0 = \nu^{-1}(132.8, 243.5, 354.1,...)$
km. 

This means that for $\nu \sim 1$ Hz, the Alfv\'{e}n waves generated in 10 flux tubes with diameters $\sim 10-20$
km, distributed over a region with a characteristic diameter $\sim 100-400$ km carry a total flux $\sim 1-5 \times 10^8$ ergs-cm$^{-2}$-sec$^{-1}$. This diameter is $\sim$ the width of inter-granular lanes, in which the kilogauss strength
flux tubes are observed to be concentrated in network and IN.

\subsection*{7.4. Photospheric Resistive Heating Flux Driven by the Alfv\'{e}n Waves}

In this section it is shown that the photospheric heating flux driven by the Alfv\'{e}n waves in the flux tubes is consistent with the values of $F_{Ph}$ in figure 3 if, as in \S 7.3, the flux tubes have a filling factor $\sim 4 \%$, and $F_{Ph}$ is
assumed to be an average over areas with diameters $\sim 100 - 400$ km.  

The period average of the resistive heating rate per unit volume driven by the waves is $\langle q \rangle = 
\eta_\parallel \langle J^2 \rangle(kR)=(\eta_\parallel/2)(c k |b_{\theta 1}|/4 \pi)^2 (J_0^2(kR)+J_1^2(kR))$. Here $J$ is the magnitude of the current density of the modes. Averaging over a flux tube area as in \S 7.3 gives $\langle q 
\rangle_{avg}= (\eta_\parallel/8)(c \nu |b_{\theta 1}|/V_A)^2 (0.3244, 0.1802, 0.1248,...)$. 

The ideal gas pressure scale height $L = k_B T/(1.3m_p g)$. Within a distance $2L$ above the
photosphere, which lies $\sim 10^2$ km below the lower chromosphere, $V_A$ varies slowly, and the product $\eta_\parallel |b_{\theta 1}|^2$ is expected to vary as $\exp(-2 z/L)/n_e \sim \exp(-z/L)$, which has an average over $2L$ of 0.4323. Then the average heating flux due to resistive heating within the height range $0 \leq z \leq 2 L$ is $F \sim (0.4323)(2L)\langle q 
\rangle_{avg}$, with $\langle q \rangle_{avg}$ evaluated at the photosphere. Since $L \sim 150$ km, $F \sim  
|b_{\theta 1}(10^2 \; \mbox{G})|^2 \nu^2 (1.9 \times 10^7, 1.1 \times 10^7, 7.3 \times 10^6,... )$ 
ergs-cm$^{-2}$-sec$^{-1}$. Taking the ratio of these values with those of $\langle S_z \rangle_{avg}$ in \S 7.3 implies that $\sim \nu^2 (43-45) \%$ of the Poynting flux is thermalized in the photosphere by electron current dissipation. 
This is consistent with the ratio $F_{Ph}/S_{Ph}$ computed from figure 3.

Multiplying these values of $F$ by 10, following the procedure applied in \S 7.3 to $\langle S_z \rangle_{avg}$, gives values comparable to those of $F_{Ph}$ in figure 3. 

\section*{8. Conclusions}

If linear, photospherically driven Alfv\'{e}nic oscillations are an important driver of resistive heating of the chromospheric, they are effective only in regions with $B_z \gtrsim 700$ G, and for $\nu \sim 10^2 - 10^3$ mHz if their magnetic field amplitude is limited to $\sim 0.01 - 0.1 B_z$, with lower amplitudes corresponding to larger $B_z$. The resistive heating is due to dissipation of ion Pedersen currents. Heavy ions dominate the Pedersen current near the height of the local temperature minimum. Protons dominate the current beginning $\sim 200$ km above this height.
Most of this heating occurs in the lower chromosphere consistent with FAL and Anderson \& Athay (1989). These results are consistent with model results in De Pontieu, Martens \& Hudson (2001) and Leake, Arber \& Khodachenko (2005), although the latter paper does not consider heating in the lower chromosphere or photosphere, and the former paper does not consider the important role of electron current dissipation in the photosphere.

The resistive heating rate in the photosphere is due to dissipation of electron currents, and exceeds the heating rate in
the chromosphere. Electron current dissipation in the photosphere limits the Poynting flux into the overlying atmosphere, limiting the chromospheric heating flux $F_{Ch}$. $F_{Ch}$ increases with $\nu$ until thermalization of Poynting flux by electron current dissipation becomes so large that $F_{Ch}$ decreases with any further increase in $\nu$. Electron current dissipation in the photosphere acts as a high frequency filter on the Poynting flux. For the parameter ranges considered in this paper, this filtering effect limits $F_{Ch}$ to $\lesssim 6 \times 10^7$ ergs-cm$^{-2}$-sec$^{-1}$. Although $F_{Ch}$ can be increased by increasing the perturbation amplitude at the photosphere, this behavior suggests that upward Poynting fluxes 1-2 orders of magnitude greater than the observed $F_{Ch}$ exist in the photosphere, but that electron current dissipation in the photosphere sets an upper bound on $F_{Ch}$ of $\sim 10^7 - 10^8$ ergs-cm$^{-2}$-sec$^{-1}$ when the photospheric Poynting flux is generated by linear Alfv\'{e}nic oscillations.

The spatial resolution needed to resolve these oscillations increases from $\sim 10$ meters in the photosphere to 
$\sim 10$ km in the upper chromosphere. The temporal resolution needed to resolve the oscillation frequencies of 
$\sim 10^2-10^3$ mHz, corresponding to periods of $1 - 10$ s, is $\sim 0.25 - 2.5$ s.

A normal mode analysis shows that Alfv\'{e}n waves with $\nu \sim 1$ Hz can be generated in
vertical kilogauss strength flux tubes in the photosphere with diameters $d \sim 10-20$ km. These waves might be excited by
resistive tearing instabilities associated with magnetic reconnection occurring near or at the boundary of the flux tube
on scales $\sim 0.01 d \sim 100$ m. Observations suggest the existence of such small diameter flux tubes, but cannot yet resolve excitation mechanisms on scales of 100 m. 

The collective results of this and earlier models of chromospheric heating by resistive dissipation lead to the conclusion that horizontal localization of the magnetic field on scales $\sim 10-10^3$ km in the photosphere and lower chromosphere are needed to generate significant heating, whether by quasi-steady convection driven heating or by MHD waves. These spatial scales are consistent with those determined from the highest resolution magnetic field observations, such as those cited in \S 1. The relative importance of quasi-steady convection driven heating, and wave driven heating
is not yet clear, but models show they can both drive the required heating rate under conditions consistent
with observations.

Nonlinear ideal MHD simulations suggest coupling between different types of waves near the $\beta=1$ surface, which is probably located in the lower chromosphere (Rosenthal et al. 2002, Bogdan et al. 2003, Goodman \& Kazeminezhad 2010, \S 5). This is expected to affect the frequency spectrum and degree of horizontal localization of the waves, thereby affecting the resistive heating rate they drive. A linear wave model, such as the one presented here, can be extended to second or higher order in the perturbation. Second order amplitudes are determined by source terms
that are quadratic in the known first order amplitudes. Higher order perturbations are determined by the known lower order perturbations. This is a simple way of estimating nonlinear effects such as wave coupling and harmonic generation.

\section*{9. Further Discussion}

The earlier models of chromospheric heating by Pedersen current dissipation driven by Alfv\'{e}nic waves mentioned in 
\S 1 are discussed in detail in this section so the model and results presented here can be understood in the context of prior work. The models use different approaches and approximations. Each model provides valuable insight. It is useful
to have a comprehensive picture showing how each model contributes to testing the viability of this heating mechanism, and indicating how more accurate models can be developed.  

Osterbrock (1961) considers chromospheric heating by dissipation of linear MHD waves, and MHD shock waves. The effect of ion-neutral collisions is taken into account in calculating the linear wave damping lengths due to resistivity and a scalar viscosity. The Hall conductivity is not considered, and heating rates are not computed. Background magnetic field
strengths $B$ of $0.5, 2$, and $50$ G, and wave frequencies $\nu \ll 50$ mHz with characteristic values $\sim 10-12$ mHz are considered. The field in plage regions is assumed to be 50 G.
The corresponding damping lengths of the linear waves are found to be too large to allow the waves to
generate significant heating in the chromosphere. It is stated that although the accurate way to compute resistive and
viscous heating rates due to shock waves is to develop a sufficiently accurate model for the structure of the shock layer, a simpler method based on jump conditions across ideal MHD shocks is used to estimate shock heating rates. The
paper concludes that shock driven heating is a significant chromospheric heating mechanism. The field strengths used in Osterbrock (1961) are 1-2 orders of
magnitude smaller than currently measured values. At such relatively low field strengths the Pedersen resistivity 
$\eta_P$, which is $\propto B^2$, is $\sim 10^2 - 10^3$ times smaller than for currently measured field strengths. This tends to reduce the resistive heating rate by the same factor. The use of such unrealistically low field strengths invalidates the conclusion that linear MHD waves cannot drive significant chromospheric heating. Regarding the use of frequencies $\ll 50$ mHz in Osterbrock (1961), it is found in the present paper in \S 6.3 that even for $B \gtrsim 10^3$ G, Alfv\'{e}n wave frequencies $\sim 10^2 - 10^3$ mHz are needed to generate significant chromospheric heating, assuming
fractional perturbations of the background magnetic field $\sim 1 - 10 \%$. 

The 1.5 D MHD model developed by De Pontieu, Martens \& Hudson (2001), which extends work by De Pontieu \& Haerendel (1998), provides an estimate of the damping rate of un-driven Alfv\'{e}n waves in the chromosphere. The wave is un-driven since it decays in time. The model includes the effect of the electrical conductivity tensor derived by Mitchner \& Kruger (1973) on Aflv\'{e}n wave dissipation in the weakly ionized region of the atmosphere, consisting of the photosphere and chromosphere. The model is a standard linear plane wave model that assumes a homogeneous background state with a constant vertical magnetic field, and subjects it to a linear perturbation with the height $z$ and time $t$ dependence $\exp(i (\omega t - k z))$. This standard model cannot predict perturbation amplitudes at a given height since they are only determined up to an overall factor by the homogeneous linear differential equations that govern the perturbation. The wave damping rate $\gamma$ is determined by obtaining the dispersion relation for $\omega(k)$, and setting $\gamma$ equal to the imaginary part of $\omega$, denoted by $\omega_i$, for the solution for which $\omega_i>0$. The corresponding wave decays in time as $\exp(- \gamma t)$, while it oscillates as $\exp(i \omega_r t)$, where $\omega_r$ is the real part of $\omega$. Then for a wave period $T=2\pi/\omega_r$, the product $q \equiv \gamma T$ is a measure of the rate at which the wave is damped. The resistive heating rate is $Q_0(z) \exp(- 2 \gamma t)$, where $Q_0(z)$ is a specified heating rate at $t=0$ at a given height. Although the model assumes a homogeneous background atmosphere, it is applied at each height in several standard, inhomogeneous background atmospheres, and, when estimating height dependent heating rates in the chromosphere the model also uses a height dependent background magnetic field strength that is 1600 G at the photosphere to compute the height dependence of the elements of the conductivity tensor. This procedure, which effectively neglects the derivatives of the background state quantities at each height, is valid only when $\lambda \ll L$, or equivalently when $\nu \gg V_A(z)/L(z)$, where $\lambda$ is the wavelength, $V_A(z)$ is the background Alfv\'{e}n speed, and $L(z) \sim k_B T/(1.3 m_p g)$ is the background pressure scale height. Requiring $\nu \gtrsim 10 V_A/L$ (the requirement $\nu \gtrsim 10^2 V_A/L$ might be more appropriate since at least 4 points are necessary to begin to resolve a wavelength), and assuming $T=8000$ K and $V_A = 10$ km-sec$^{-1}$ as characteristic values for the chromosphere suggests the model is valid for $\nu \gtrsim 500$ mHz. The model predicts that Alfv\'{e}n waves experience strong damping for $\nu \gtrsim 20, 200$, and 2000 mHz in sunspot umbrae, quiet Sun, and plage regions, respectively. The lower bound on $\nu$ for umbrae is too low, and that for quiet Sun is probably too low for the neglect of the local inhomogeneity of the background state to be valid. However, the model predictions that $q$ increases with $\nu$, and that $q$, and hence the heating rate is largest in the
lower chromosphere, defined by the height range $z \sim 500 - 1000$ km above the photosphere, are consistent with the results presented here in \S 6.3, although there it is also found that above a certain frequency electron current
dissipation of the upward propagating Poynting flux in the photospheric region $z \lesssim 500$ km causes $q$ to decrease
with increasing $\nu$ for sufficiently large $\nu$.

In addition to the local, un-driven excitation of waves on time scales $\ll$ the linear wave damping time $\sim 
\gamma^{-1}$, waves may be driven in a steady or quasi-steady manner on longer time scales, for example by wave generation in the photosphere. For linear waves this is the case of a driven oscillator, for which the wave amplitudes
oscillate at the driving frequency, and do not decay in time. A dynamic equilibrium is established in which the rate at
which the wave generating process pumps energy into the wave equals the rate at which the wave loses energy by the wave
electric field driving currents that experience resistive dissipation. The model presented here considers such a dynamic equilibrium driven by Alfv\'{e}nic oscillations.

Khodachenko, Arber, Rucker \& Hanslmeier (2004) have the objective of determining the relative importance of resistive
and viscous damping of un-driven, local, linear MHD waves from the photosphere to the corona, and in prominences. The
method used to do this is to compute ratios of resistive damping times to viscous damping times. This method is limited in its ability to determine the relative importance of resistive and viscous heating for the following reason. For the case of un-driven, local, linear waves, which is the case considered in the paper, the characteristic form of the resistive and viscous heating rates are as follows. The resistive heating rate $Q_{res} \sim 
(E_1^2(0)/\eta) \exp(-t/\tau_{res}) \equiv Q_{res}(0) \exp(-t/\tau_{res})$. Here $E_1(0)$ is the perturbed electric field
at time $t=0$, $\eta$ is the background resistivity, and $\tau_{res}$ is the resistive damping time computed in the
paper. The total viscous heating rate $Q_{vis} \sim A_{\alpha \beta \gamma \delta} \partial_\alpha V_{\beta 1}(0) 
\partial_\gamma V_{\delta 1}(0) \exp(-t/\tau_{vis}) \equiv Q_{vis}(0)\exp(-t/\tau_{vis})$. Here $V_{\alpha 1}(0)$ is the 
$\alpha$ component of the perturbed CM velocity at $t=0$, the $A_{\alpha \beta \gamma \delta}$ are the background state viscosity coefficients, the sum is over repeated indices with values $x,y,z$, and $\tau_{vis}$ is the viscous damping time computed in the paper. The paper uses $\tau_{res}/\tau_{vis}$ as a measure of the relative importance of resistive and viscous heating. This comparison neglects the fact that $Q_{res}(0)$ and $Q_{vis}(0)$ may differ by orders of magnitude. The former heating rate is driven by an electric field, the latter by velocity gradients. Khodachenko, Arber, Rucker \& Hanslmeier (2004) conclude that viscous damping can be important relative to resistive damping. However, the only reliable way to estimate the relative importance of $Q_{res}({\bf x},t)$ and $Q_{vis}({\bf x},t)$ is to compute their ratio. This can only be done using a model that solves for the electric and velocity field perturbations.

Viscous heating determined by the sum of the anisotropic electron and ion viscosity tensors, and the isotropic neutral gas viscosity tensor might be important. The strong magnetization of the chromosphere decreases viscosities involving directions orthogonal to ${\bf B}$ by orders of magnitude. The viscous heating rate involves products of viscosities and spatial derivatives of components of ${\bf V}$, so the effect of strong magnetization on these derivatives must be determined. Similarly, the model does not include the compressive heating rate $-p \nabla \cdot {\bf V}$. The reason is that the background magnetic field is purely vertical. A horizontal component is needed to couple compressive MHD and acoustic modes into the model.

Leake, Arber \& Khodachenko (2005) present 1 D analytic and numerical MHD models of Alfv\'{e}n wave damping in the strongly magnetized region of the chromosphere. The models assume a background atmosphere given by the VAL C model
(Vernazza, Avrett \& Loeser 1981), the Ohm's law derived by Braginskii (1965), and a height dependent background magnetic
field strength $B$. The assumption of strong magnetization restricts the validity of the models to the region $z \geq 
750$ km. The damping mechanism includes Pedersen current dissipation, and omits the Hall conductivity based on the
assumption of strong magnetization\footnote{In linear theory a small error is incurred by omitting the Hall conductivity 
$\sigma_H$ in the strongly magnetized chromosphere. The reason is that it couples orthogonal components of ${\bf B}$
through Faraday's law, which can increase the resistive heating rate. For example, in the case of Alfv\'{e}n waves, 
$\sigma_H$ couples $B_x$ and $B_y$, and causes them to be essentially equal. This doubles the heating rate over what it
is in the absence of $\sigma_H$ since either $B_x$ or $B_y$ can be chosen to be zero in this case. In nonlinear theory the effect of $\sigma_H$ is stronger since it can lead to the generation of frequency
components, with a significant amplitude, in the power spectrum of ${\bf B}$ at frequencies higher than the driving
frequency. This corresponds to the generation of currents on smaller spatial scales and to increased resistive dissipation.} The photosphere and lower chromosphere, which lie in the region $0 \leq z \leq 1000$ km of VAL C, are mostly excluded from the model. About $2/3$ of the net radiative loss from the chromosphere is emitted from the lower chromosphere, which lies in the height range $500 \lesssim z(\mbox{km}) \lesssim 10^3$ (Anderson \& Athay 1989; FAL). Most chromospheric heating occurs in the lower chromosphere. The wave amplitude $A$ in the analytic model has the
form $A(z,t)=A_0(z_0,0) \exp(i(k(z-z_0)-\omega_r t)) \exp(-\gamma t)$, where the damping rate $\gamma \propto 
\omega_r^2$. Using this form, the model estimates the relative change in a wave amplitude with height as $|A/A_0| = 
\exp(-\gamma t) \sim \exp(-\int_{z_0}^z (\gamma(z)/V_A(z))\;dz)$, where $V_A(z)$ is the background Alfv\'{e}n speed, and
it is assumed that $dz=V_A(z) \; dt$ in the sense of a wavepacket propagating with the local speed $V_A(z)$. Although this model cannot predict wave amplitudes, the predicted relative variation of amplitude with height is compared with results of 1.5 D MHD simulations of Alfv\'{e}n waves driven by a harmonic, horizontal velocity perturbation at $z=750$ km,
where $B \sim 10^3$ G. The simulations use a resolution of 1.5 km. The overall agreement between the relative damping of the Poynting flux predicted by the linear wave model and by the MHD simulation as a function of driving frequency for $z \geq 10^3$ km is good for the frequency range $0 \leq \nu \leq 600$ mHz. The level of agreement increases with frequency, and hence with damping rate. The values of the wave's magnetic field and velocity amplitudes, and the background magnetic field strength at $z=10^3$ km for the simulation are given for a typical case in figures 2 and 6 of the paper as $\delta B \sim 0.8$ G, $\delta V \sim 13.03$ m-sec$^{-1}$, and $B \sim 700$ G, where $\nu =70$ mHz. Then the period averaged upward Poynting flux at this height is $S_z = B \delta V \delta B/4 \pi \sim 5.8 \times 10^4$ ergs-cm$^{-2}$-sec$^{-1}$. This is $10^2-10^3$ times too small to drive the chromospheric NRL for $z \gtrsim 10^3$ km, assuming all of $S_z$ is converted into thermal energy in the chromosphere. In the linear approximation, $\delta B \propto \delta V$, so increasing $\delta V$ by a factor $f$ increases $S_z$ by a factor $f^2$. Then wave amplitudes $\gtrsim 10$ times larger than those used in the simulation are necessary to drive the NRL for $z \gtrsim 10^3$ km. Although the simulation is
driven at $z=750$ km, there is no discussion of damping below $z=1000$ km. This is puzzling since most
chromospheric heating occurs below $z=10^3$ km. In this context, it is stated in the paper that a typical driving
velocity amplitude $\delta V = 600$ m-sec$^{-1}$. This is 46 times larger than the velocity amplitude in figure 6 at 
$z=1000$ km. Using linear theory, $\delta B \sim (4 \pi \rho)^{1/2} \delta V$, where $\rho$ is the background density.
Assuming $\delta V=600$ m-sec$^{-1}$ at $z=750$ km, using $B(750 \; \mbox{km})=1000$ G from figure 5, and $B(10^3 \; 
\mbox{km})=700$ G, and using equation (25) of the paper to determine $\rho (750 \; \mbox{km})$ gives $\delta B (750 \; 
\mbox{km}) \sim 82$ G. Then $S_z(750 \; \mbox{km}) \sim 3.9 \times 10^8$ ergs-cm$^{-2}$-sec$^{-1}$. This is more than
enough to balance the NRL of the entire chromosphere. However, if this is indeed the order of $S_z(750 \; \mbox{km})$ in
the simulation then virtually all of the wave energy is dissipated in the height range $750 \leq z \leq 1000$ km.
It is also puzzling that no resistive heating rates versus height are presented since the simulation determines all quantities needed to compute them. Then the main result of the paper must be taken to be that linear Alfv\'{e}n waves associated with a Poynting flux at $z=10^3$ km that is 2-3 orders of magnitude smaller than the chromospheric NRL are strongly damped in the chromosphere above $z=1000$ km by Pedersen current dissipation for $\nu \gtrsim$ several hundred mHz, but are largely undamped at much lower frequencies. The result that damping increases with frequency is also found in De Pontieu, Martens \& Hudson (2001), and in the model presented here, although here heating rates are computed from the photosphere into the lower corona, and, as mentioned above, in \S 6.3 it is shown that resistive dissipation of electron currents in the photosphere can significantly reduce the Poynting flux into the overlying atmosphere, causing the chromospheric heating flux to decrease.

KG06 use a 1.5 D nonlinear MHD simulation to compute the heating due to dissipation of Alfv\'{e}n waves in a background
FAL atmosphere with a constant $B_z= 25$ G, and a conductivity tensor evaluated using a height dependent magnetic field
strength equal to 1500 G at the photosphere.\footnote{KG06 also includes an analysis of the dispersion relations for
linear Alfv\'{e}n and magnetoacoustic waves modified by the presence of the anisotropic conductivity tensor with Hall,
Pedersen, and Spitzer conductivities.} The numerical resolution $\Delta= 10$ km. Numerical dissipation is shown to be
insignificant, so the equations of the model are solved accurately. Waves are driven at $z=1000$ km by an oscillating
horizontal magnetic field with an amplitude of 5 G. The photosphere and lower chromosphere are omitted from the model.
The Hall terms in Faraday's law and the energy equation are artificially increased by a factor of $10^3$, which is the
ratio of $\Delta$ to the Hall length scale, in an attempt to model the small scale effects of the Hall terms (see \S 3.4 of KG06). The duration of the simulation is 28 wave periods. The wavelength and frequency of the driver are $\lambda=170$ km and $\nu=33$ mHz. About $83.5 \%$ of the work done by the driver is converted into thermal energy. The waves are essentially completely damped within a distance $\sim \lambda$ above $z=1000$ km, and generate a period averaged resistive heating rate $Q \sim 0.05$ ergs-cm$^{-3}$-sec$^{-1}$, corresponding to a local heating flux $\sim Q \lambda = 8.5 \times 10^5$ ergs-cm$^{-2}$-sec$^{-1}$ generated over a height range of $\lambda$. This rapid damping at relatively low frequency is probably due to the use of a $V_A$ computed using the constant BG field of 25 G, rather than the height
dependent $B$ used to compute the conductivity tensor. Using $B$ to compute $V_A$ increases it by a factor that decreases from 5.2 at $z=10^3$ km to $\sim 1$ at $z=2000$ km. This increase in $V_A$ causes the waves to propagate faster, dissipate
energy over a larger height range, and possibly experience much less total dissipation since they move more quickly 
through the chromosphere.

The 1 D models of De Pontieu, Martens \& Hudson (2001), Khodachenko, Arber, Rucker \& Hanslmeier (2004), Leake, Arber \& Khodachenko (2005), and KG06 use a height dependent background magnetic field strength to compute the electrical
conductivity tensor in the chromosphere for various cases involving Alfv\'{e}nic oscillations. This is a reasonable first
approximation for inserting a height dependent conductivity tensor into a 1 D model. However, the approximation has two
significant deficiencies that are not discussed in these papers. They are as follows: (1) A height dependent magnetic
field strength in a 1 D model is not consistent with Alfv\'{e}n waves. The only field components that can vary with $z$
in a 1D model are horizontal components. If a horizontal field is present then magnetoacoustic waves are coupled into the
model. The solutions then become magnetoacoustic waves if $B_z=0$, or waves that represent a coupling between Alfv\'{e}n
and magnetoacoustic waves if $B_z \neq 0$. (2) The presence of a height dependent background magnetic field in a model
implies the background state has a nonzero resistive heating rate since ${\bf J} \neq 0$, except in the special case of a
potential field. The significance of the heating rate due to linear wave damping in such a model can only be determined
by comparing it with the background state heating rate, which might be $\gtrsim$ the wave driven heating rate. Since the background state is a steady state, its heating must be driven by a convection electric field.

The model presented here assumes a constant, vertical background magnetic field extending from the photosphere into the lower corona. This is consistent with the 1 D approximation, and allows for pure Alfv\'{e}n wave solutions, but does not model the expected decrease of $B$ with increasing height. The decrease of $B$ with height is expected to reduce the
wave driven chromospheric heating flux since the background $\eta_P \propto B^2$ in the chromosphere, but it is also expected to increase the background heating rate by introducing a nonzero background current density. It is expected that the more rapidly $B$ decreases with increasing height, the greater the reduction in the wave driven heating flux, and the larger the background heating rate. The effect of a height dependent background field can be properly estimated only by
including a height dependent horizontal magnetic field in the model, or by using a multi-dimensional model, which allows
a height dependent vertical field. The heating rates predicted by the model presented here are estimates of heating rates in strong field regions where the field is mainly vertical. For sunspot umbrae, observations suggest the model is valid up to the lower corona since in umbral regions $B$ may decrease by only a factor $\sim 2$ between the photosphere and lower corona, with a mean rate of decrease over a height of 2000 km $\sim 0.3-0.6$ G-km$^{-1}$(Solanki 2003 \S 3). 

\begin{acknowledgements}
This work was supported by grant ATM-0650443 from the Solar-Terrestrial Physics Program of the National Science
Foundation to the West Virginia High Technology Consortium Foundation. Part of this work was done during a visitor
appointment at the High Altitude Observatory (HAO) of the National Center for Atmospheric Research. The author thanks the
HAO staff, especially Philip Judge and BC Low, for their generous hospitality. This research made use of NASA's Astrophysics Data System (ADS).
\end{acknowledgements}

\newpage
\figcaption[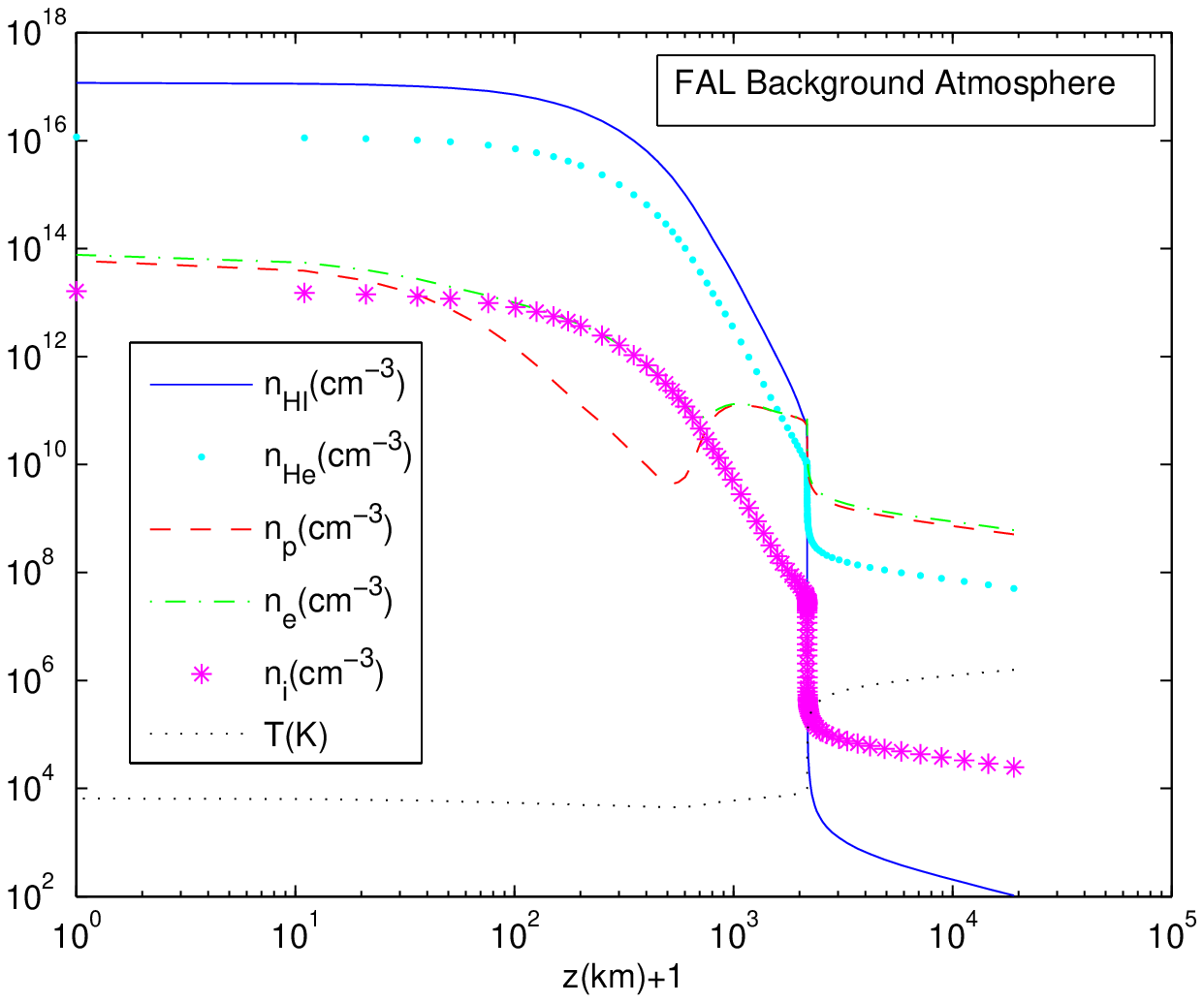]{FAL background state profiles. Shown are the temperature, densities for HI, protons and electrons,
total He density, and the total density $n_i$ of the singly charged heavy ions of C, Si, Al, Mg, Fe, Na, and Ca.}

\figcaption[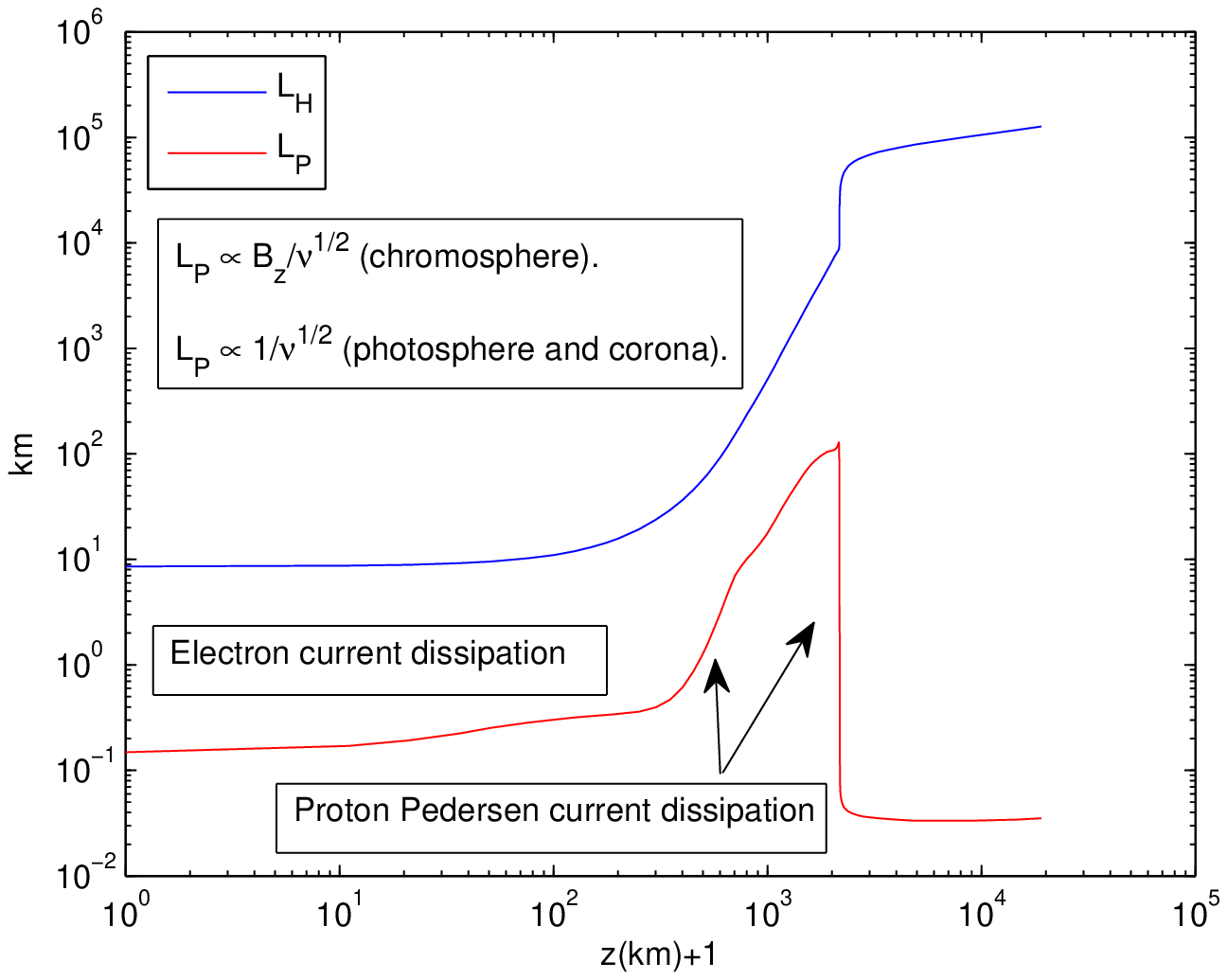]{Length scales $L_P$ and $L_H$ vs. $z$ for $B_z=10^3$ G and $\nu = 100$ mHz. The scaling of $L_P$ with
$B_z$ and $\nu$ is also indicated.}

\figcaption[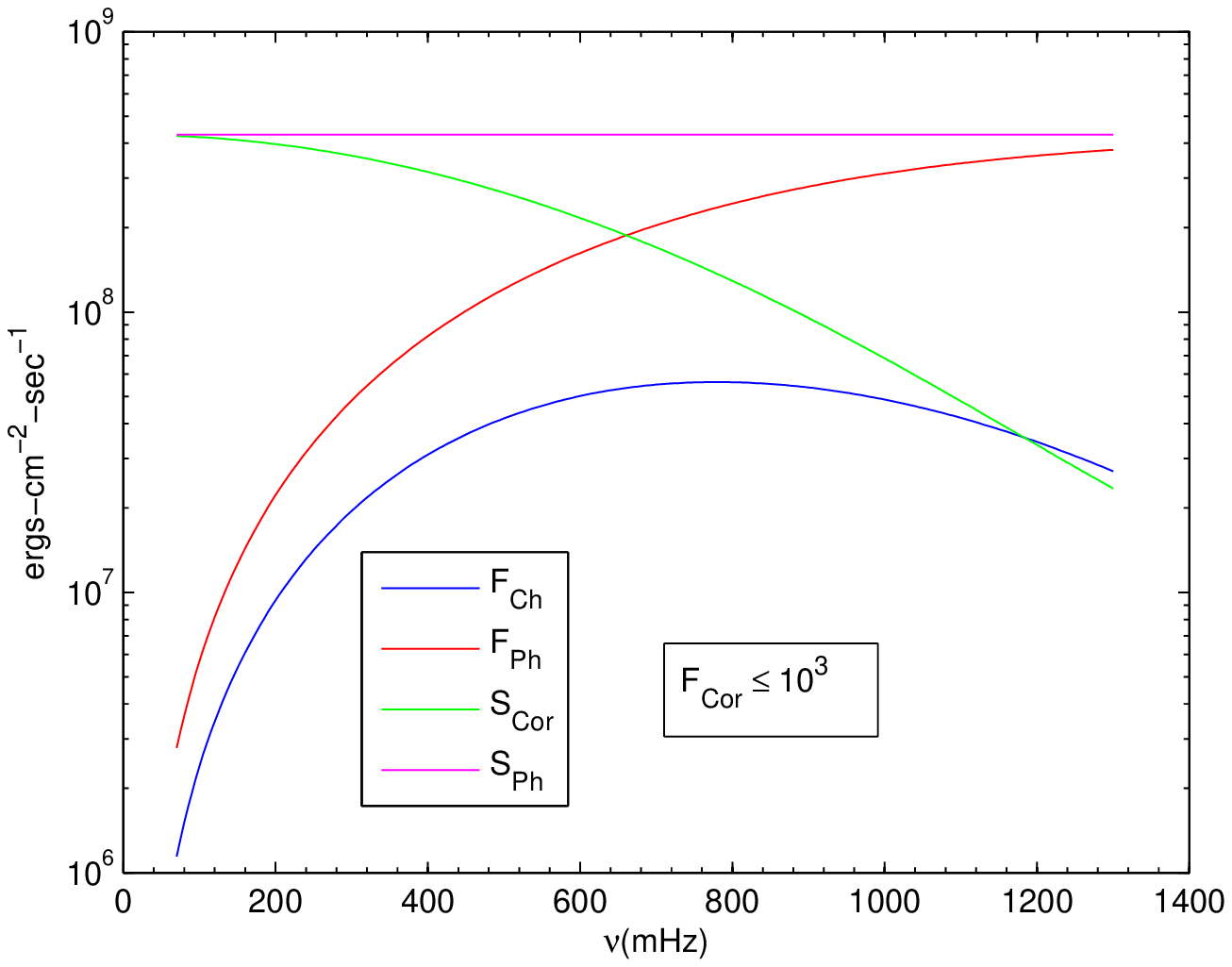]{Heating rates and fluxes vs. $\nu$ for $B_z=10^3$ G.}

\figcaption[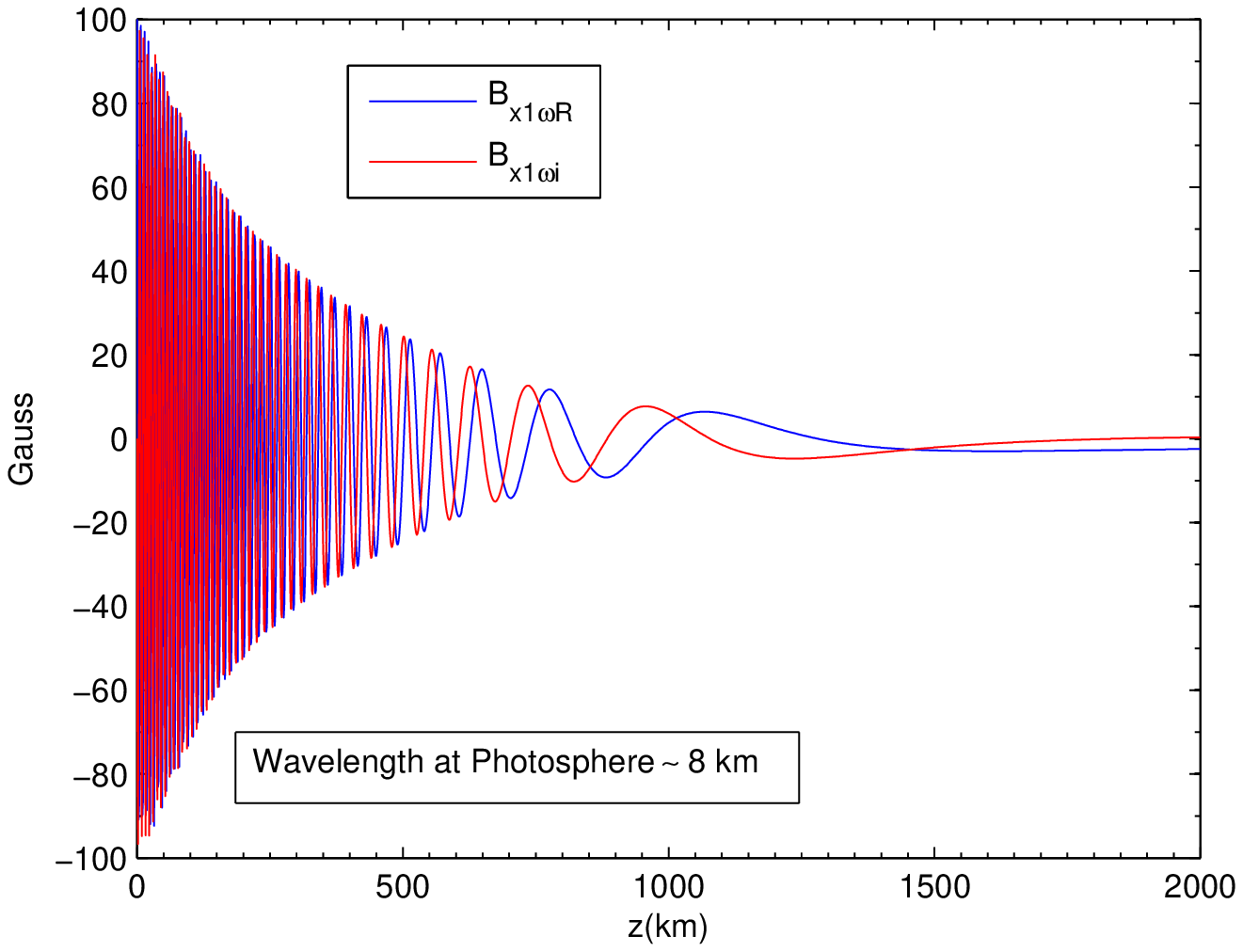]{Real and imaginary parts of the $x$ component of the magnetic field perturbation vs. $z$ for $B_z=
10^3$ G, and $\nu=780$ mHz, which is the frequency at which $F_{Ch}$ is a maximum.}

\figcaption[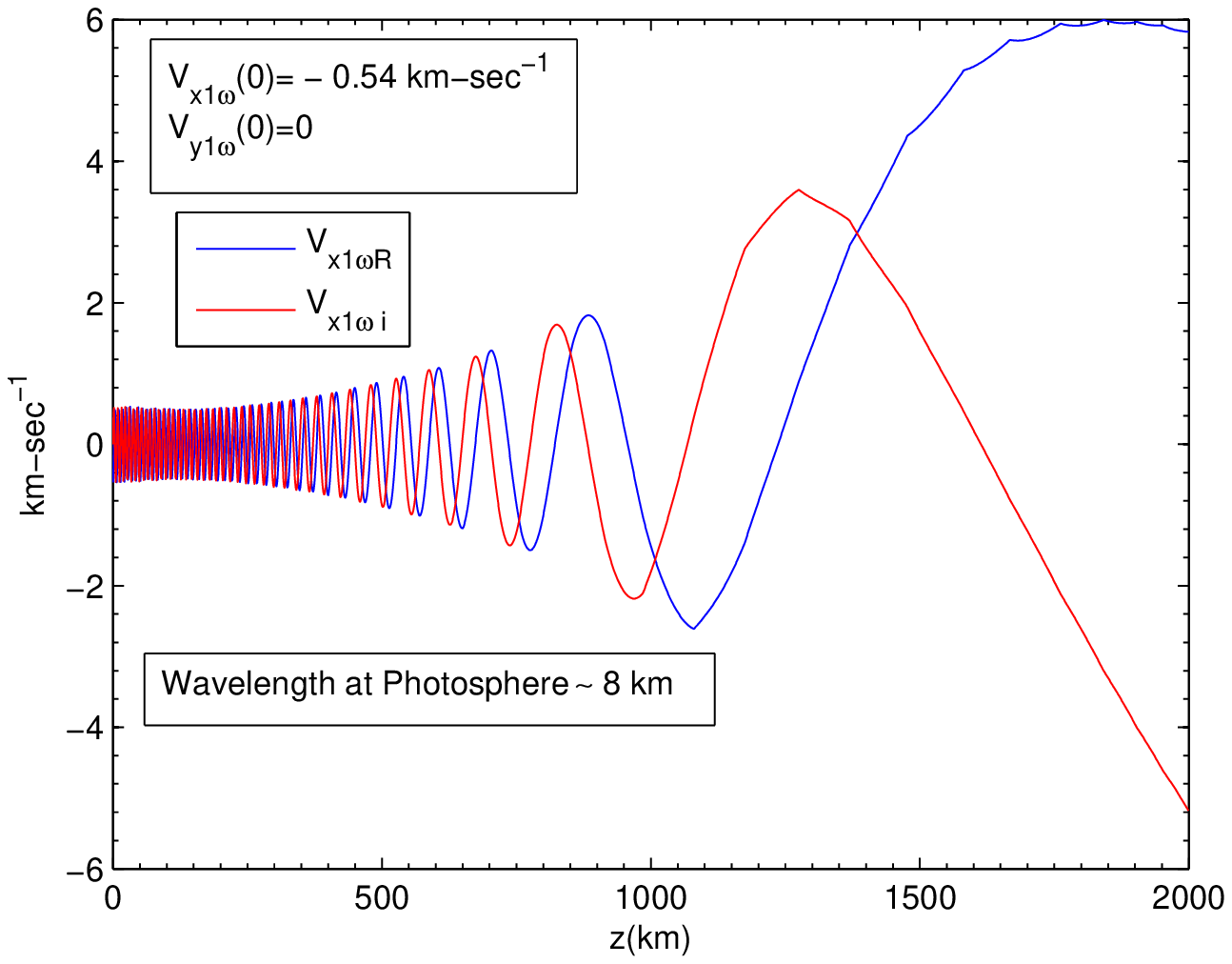]{Real and imaginary parts of the $x$ component of the velocity field perturbation vs. $z$ for $B_z=
10^3$ G, and $\nu=780$ mHz, which is the frequency at which $F_{Ch}$ is a maximum.}

\figcaption[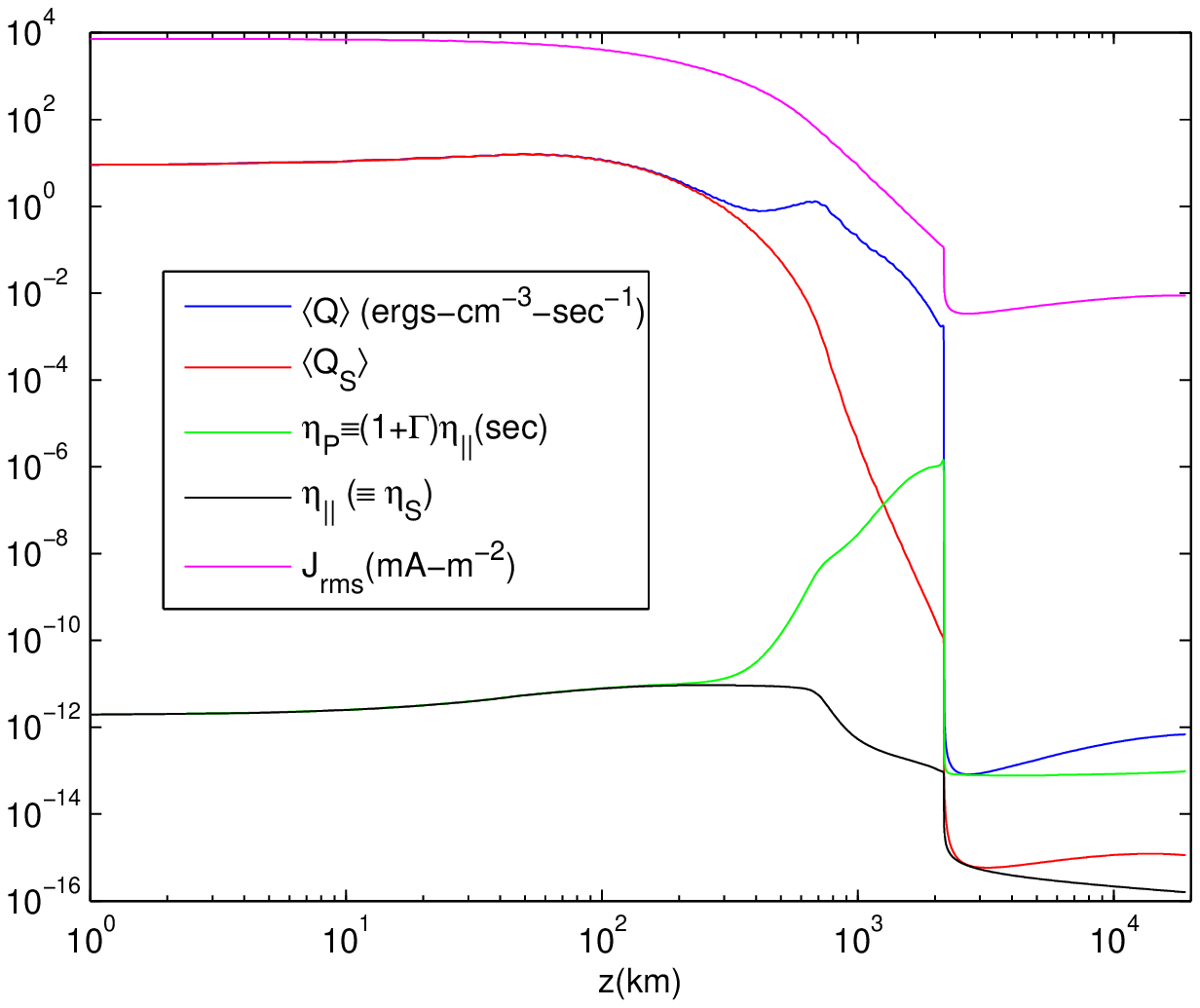]{Period averaged total and Spitzer resistive heating rates per unit volume, Pedersen and Spitzer resistivities, and rms current density vs. $z$ for $B_z=10^3$ G, and $\nu=780$ mHz, which is the frequency at which 
$F_{Ch}$ is a maximum.}

\figcaption[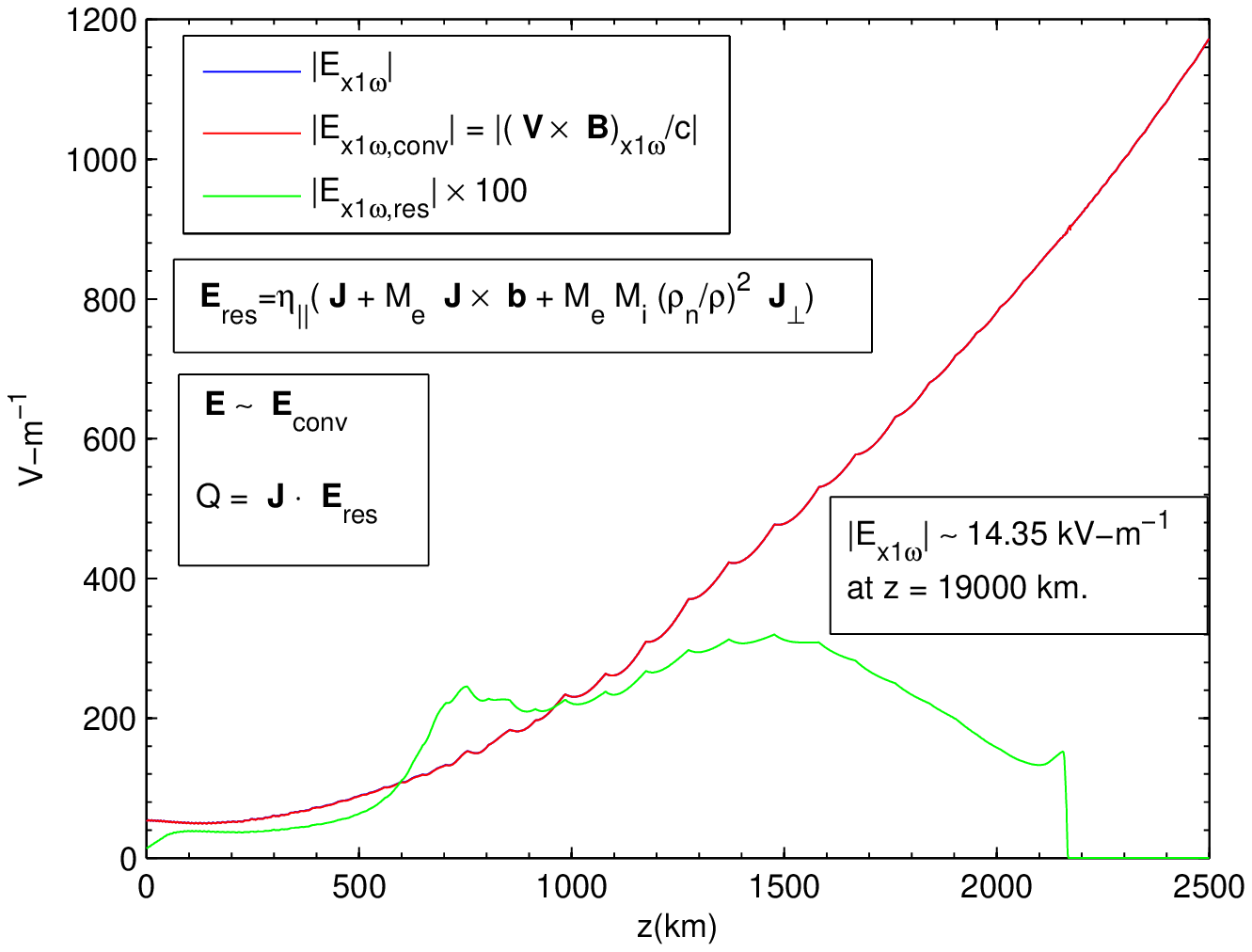]{Magnitudes of the amplitudes of the $x$ component of the electric field perturbation and its convection and resistive components vs. $z$ for $B_z=10^3$ G, and $\nu=780$ mHz, which is the frequency at which 
$F_{Ch}$ is a maximum.}

\figcaption[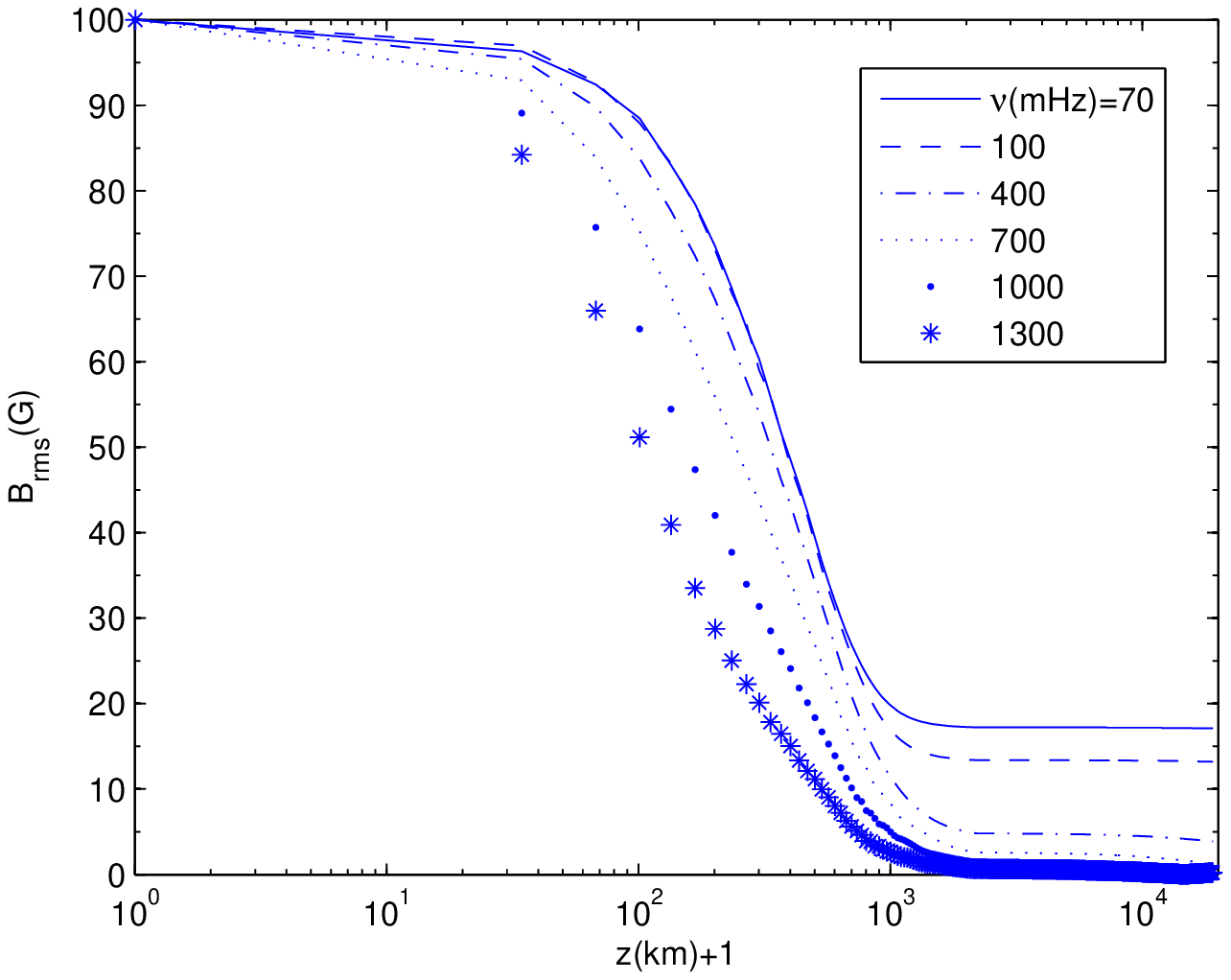]{rms of the magnetic field perturbation vs. $z$ for a range of $\nu$, and $B_z=10^3$ G.}

\figcaption[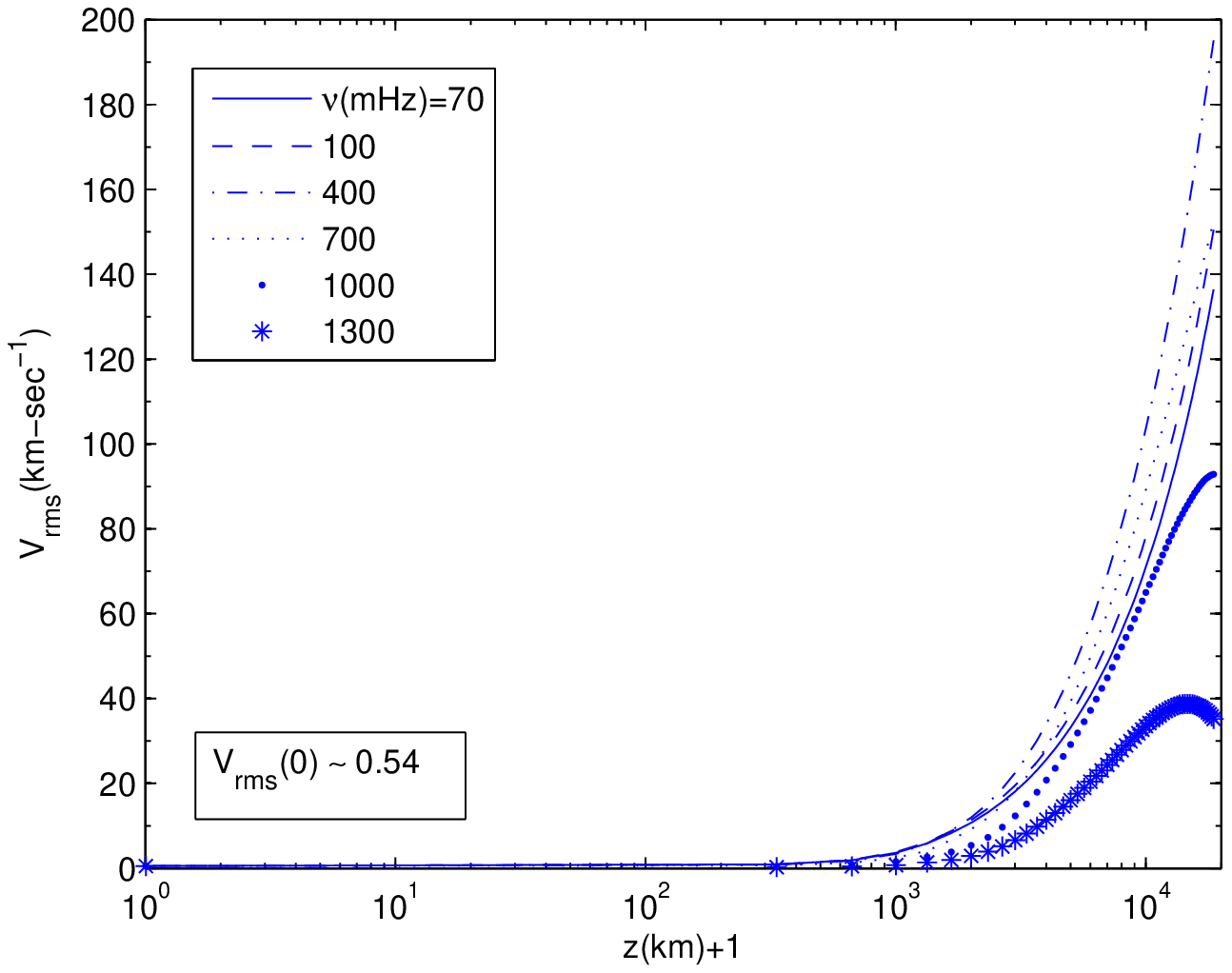]{rms of the velocity field perturbation vs. $z$ for a range of $\nu$, and $B_z=10^3$ G.}

\figcaption[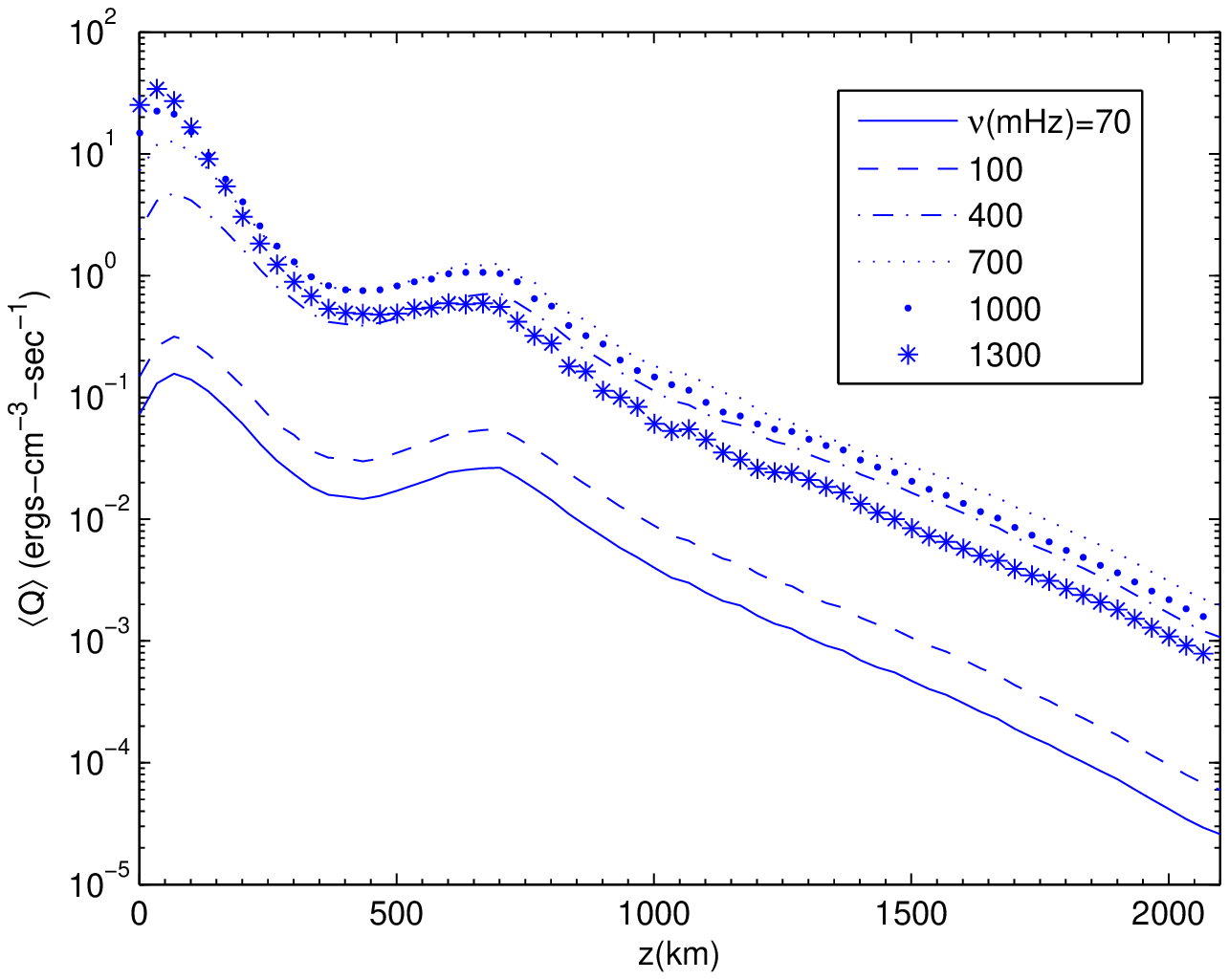]{Period averaged resistive heating rate per unit volume vs. $z$ for a range of $\nu$, and $B_z=10^3$ G.}

\figcaption[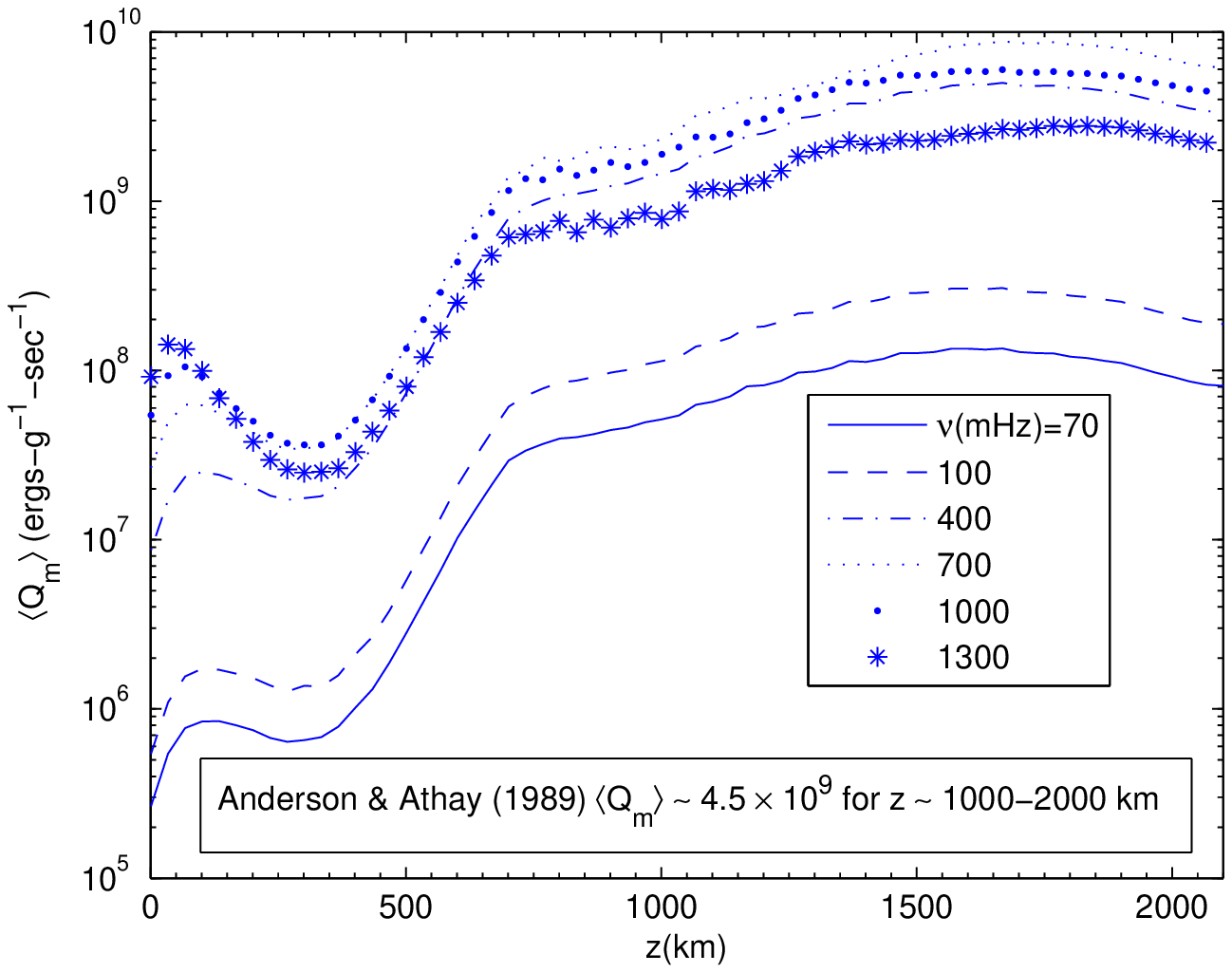]{Period averaged resistive heating rate per unit mass vs. $z$ for a range of $\nu$, and $B_z=10^3$ G.}

\figcaption[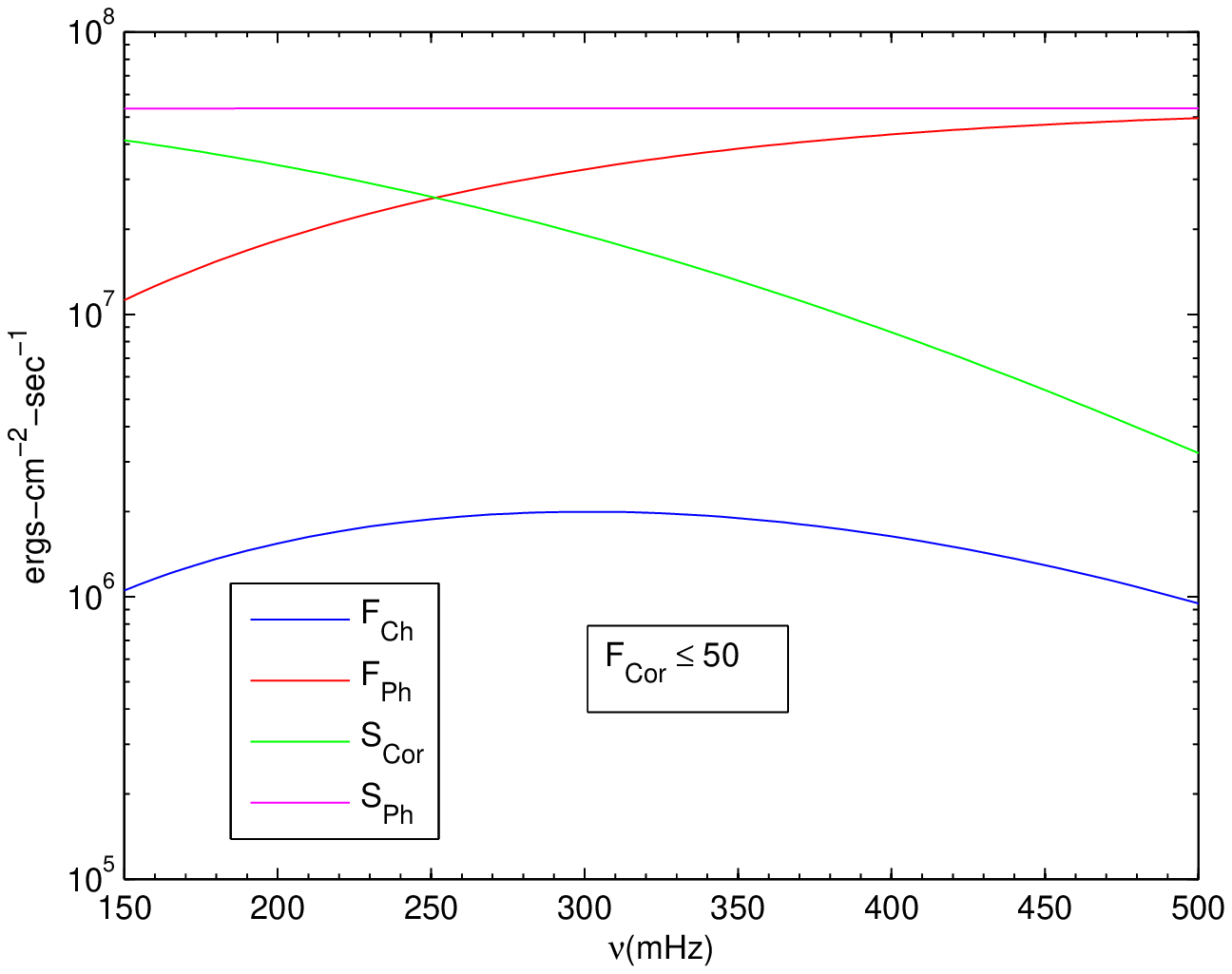]{Heating rates and fluxes vs. $\nu$ for $B_z=500$ G.}

\figcaption[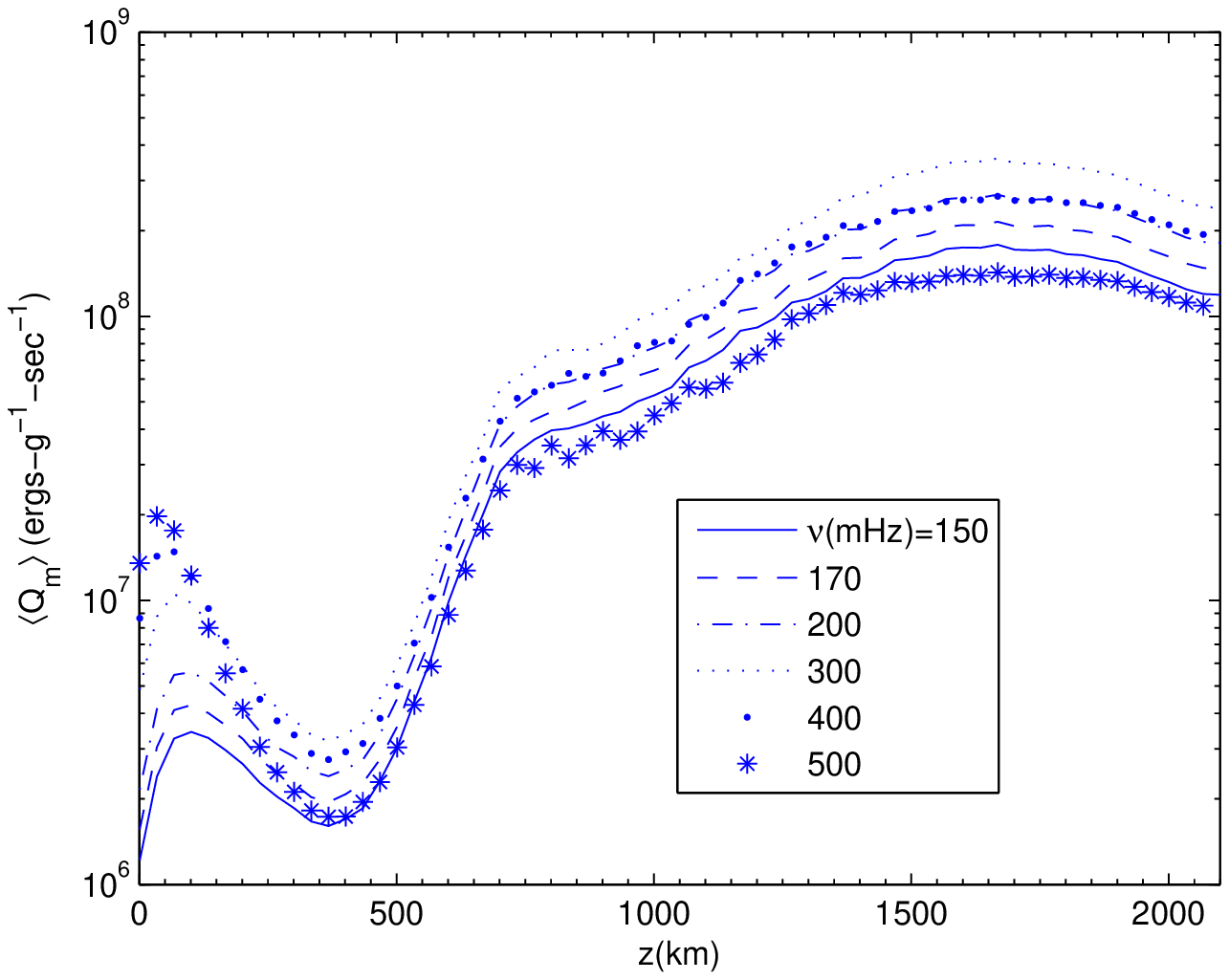]{Period averaged resistive heating rate per unit mass vs. $z$ for a range of $\nu$, and $B_z=500$ G.}

\figcaption[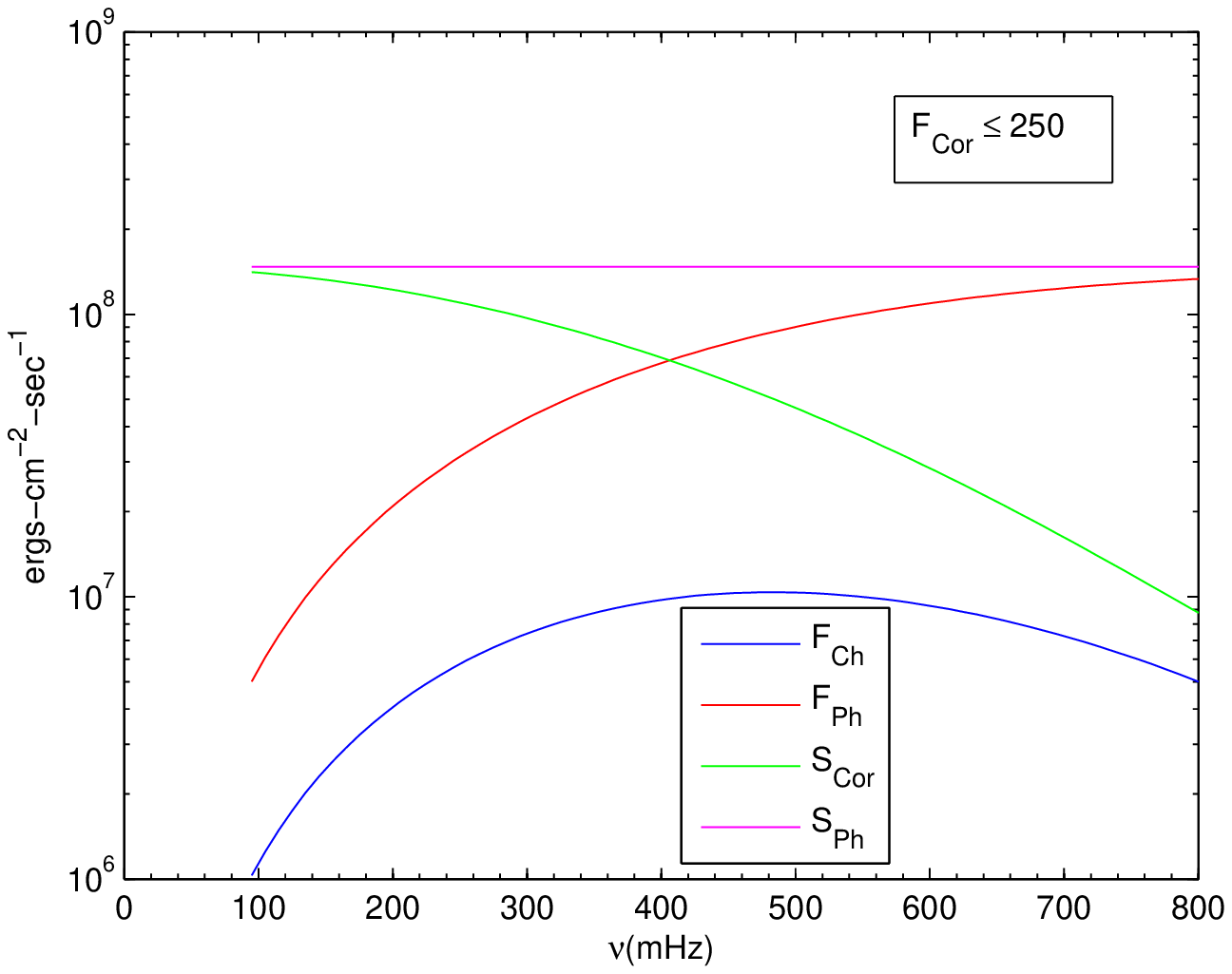]{Heating rates and fluxes vs. $\nu$ for $B_z=750$ G.}

\figcaption[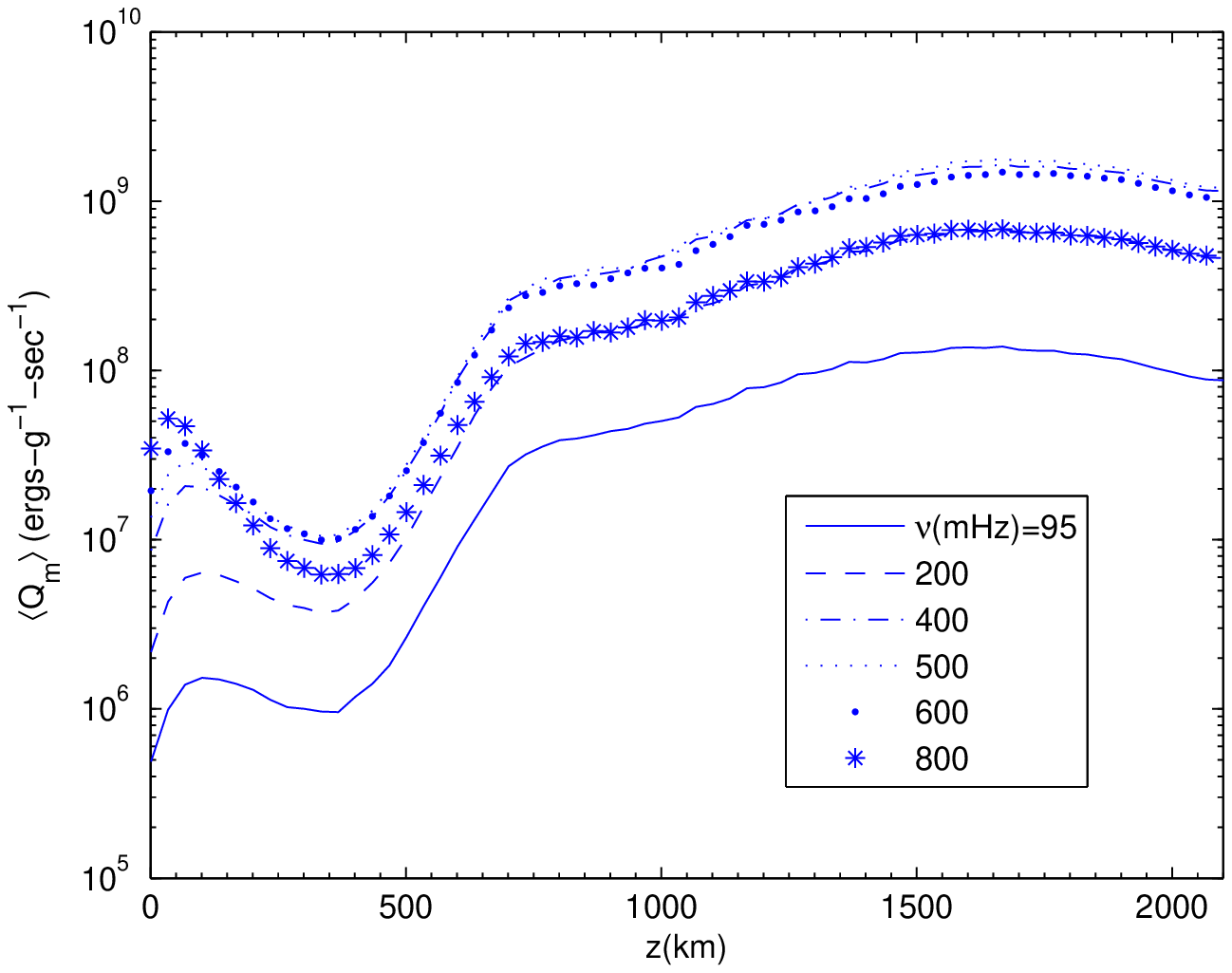]{Period averaged resistive heating rate per unit mass vs. $z$ for a range of $\nu$, and $B_z=750$ G.}

\figcaption[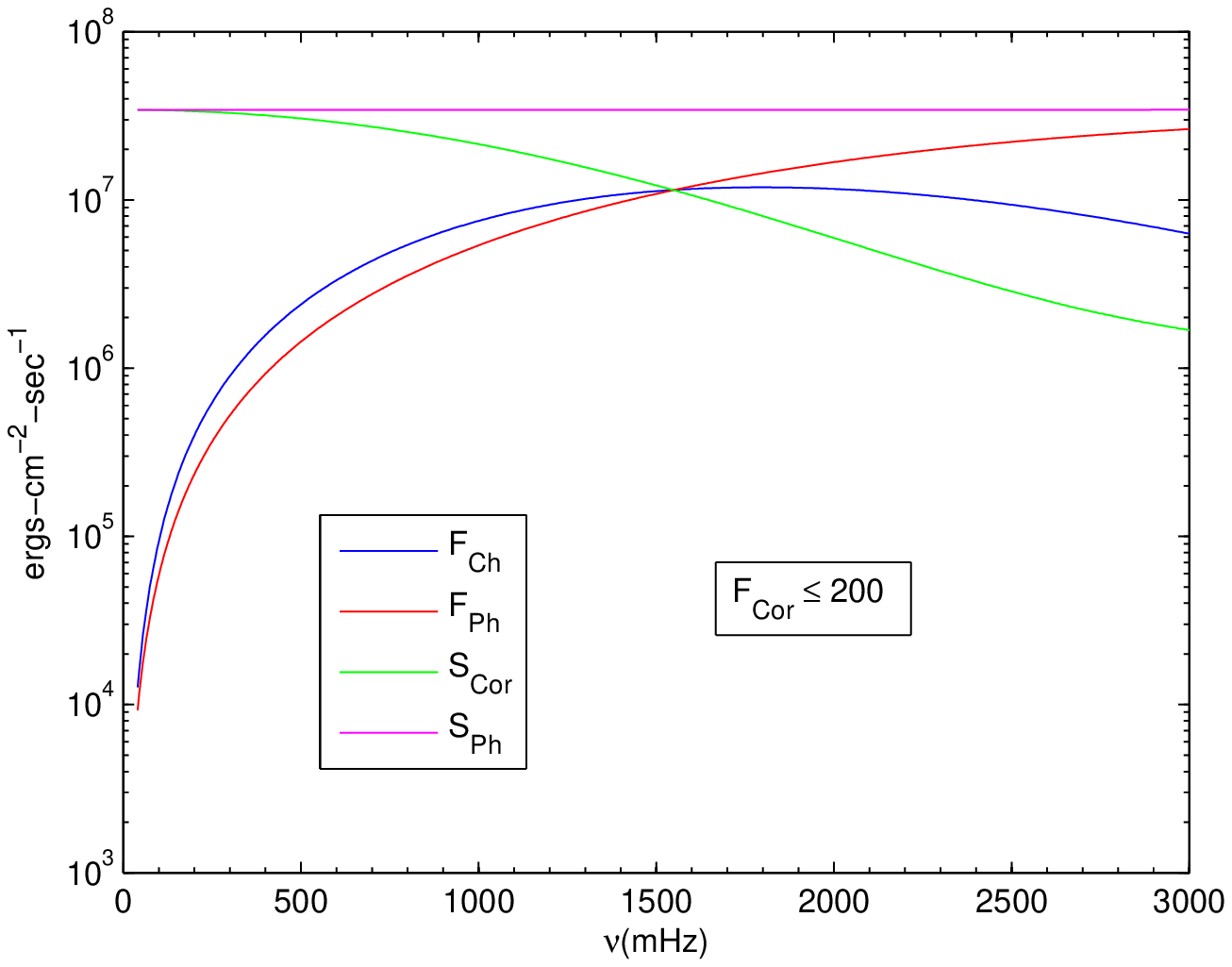]{Heating rates and fluxes vs. $\nu$ for $B_z=2000$ G.}

\figcaption[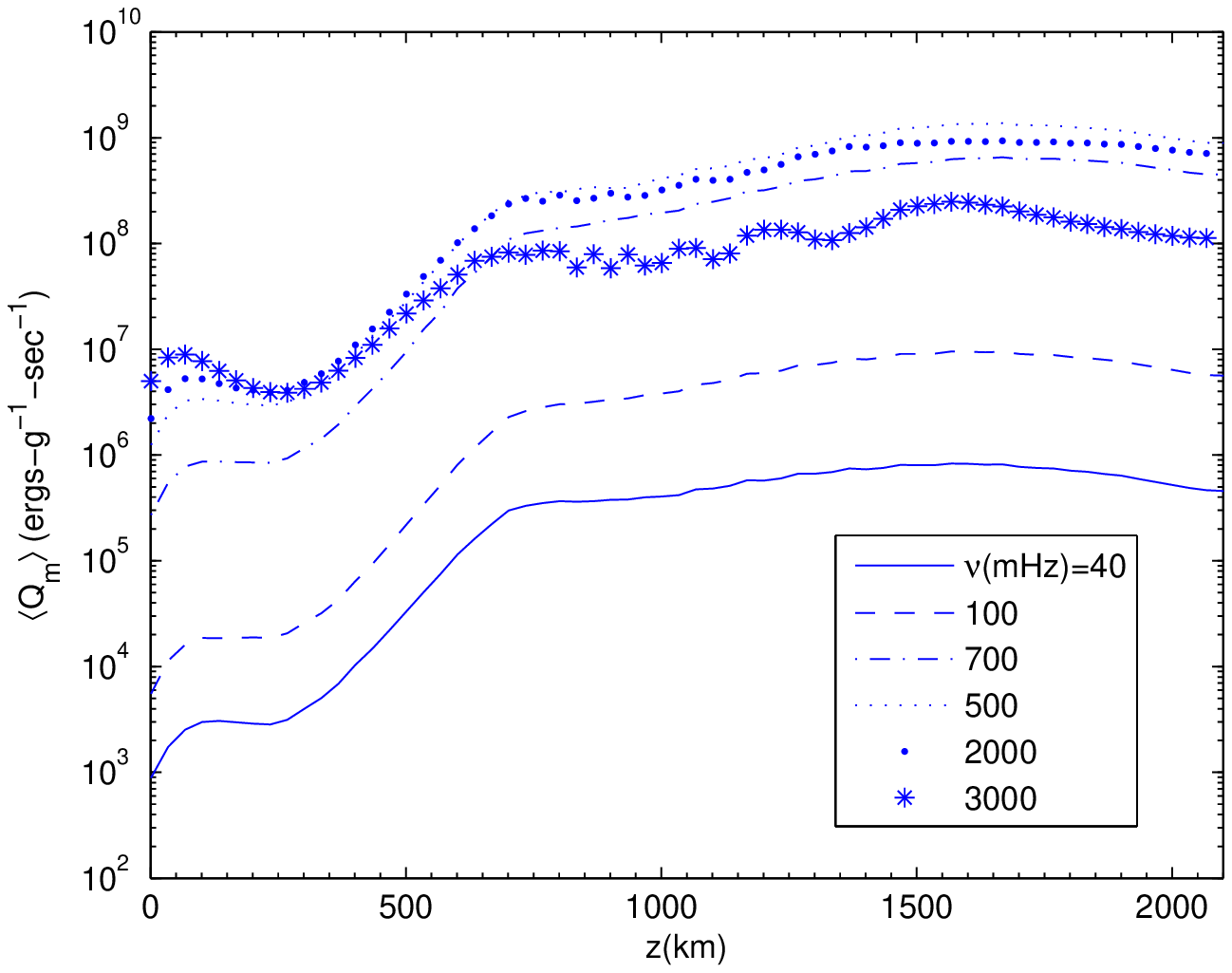]{Period averaged resistive heating rate per unit mass vs. $z$ for a range of $\nu$, and $B_z=2000$ G.}

\figcaption[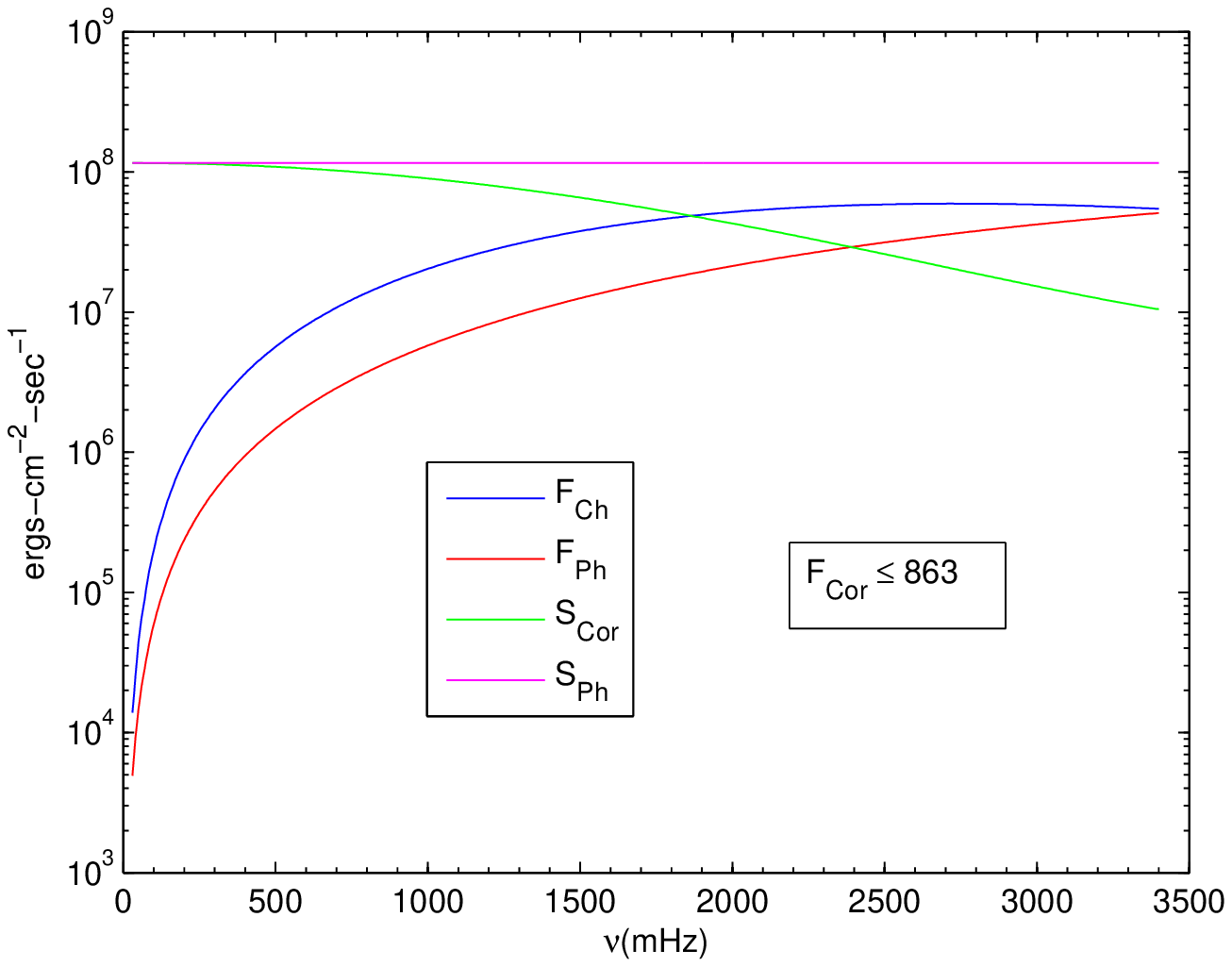]{Heating rates and fluxes vs. $\nu$ for $B_z=3000$ G.}

\figcaption[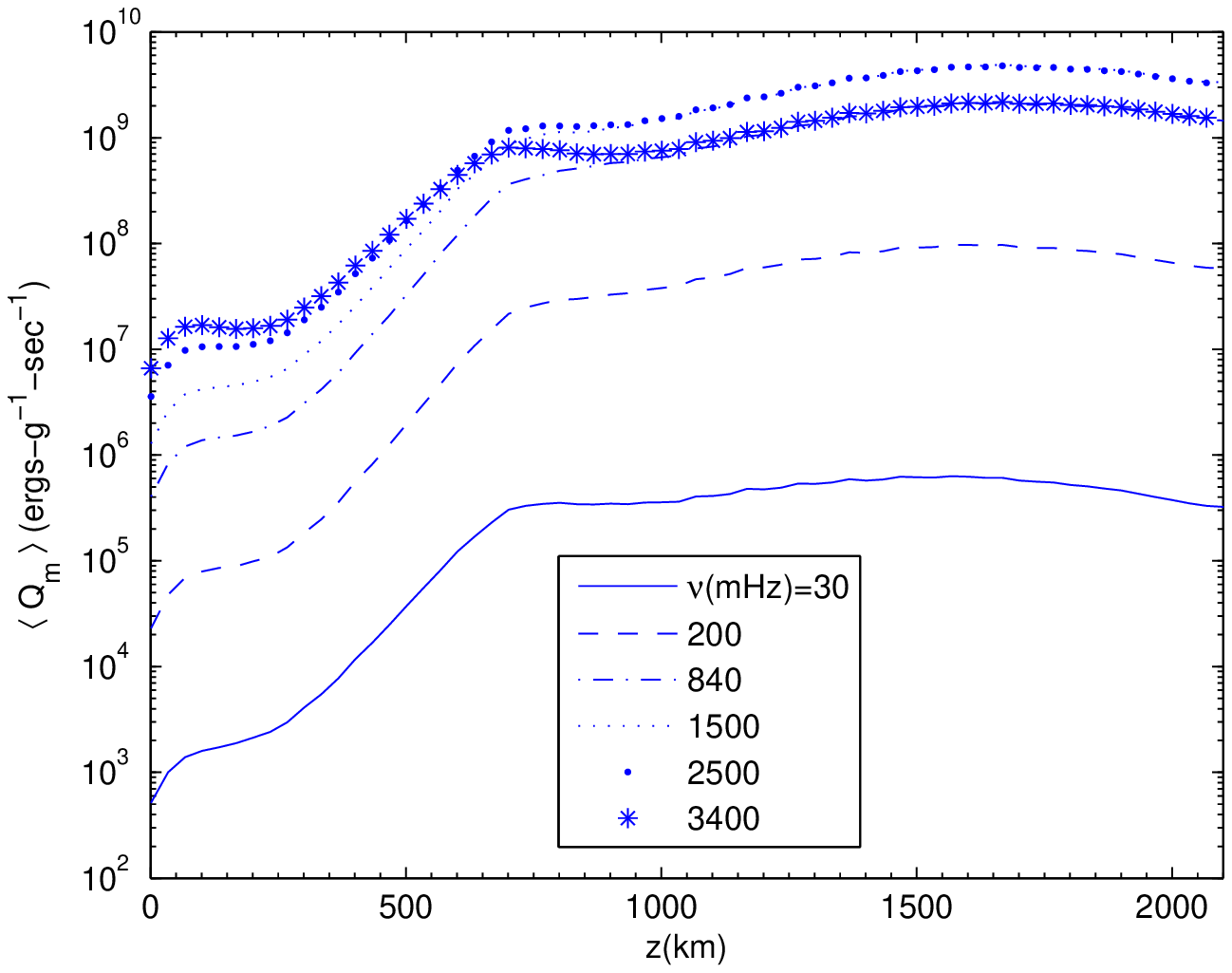]{Period averaged resistive heating rate per unit mass vs. $z$ for a range of $\nu$, and $B_z=3000$ G.}
\end{document}